\pdfoutput=1
\RequirePackage{fix-cm}
\documentclass[journal,british,twoside]{IEEEtran}

\usepackage[british]{babel}

\usepackage{cite}

\ifCLASSINFOpdf
  \usepackage[pdftex]{graphicx}
  \graphicspath{{./figures/}}
  \DeclareGraphicsExtensions{.pdf,.png,.jpeg,.jpg}
\else
\fi
\usepackage[cmex10]{amsmath}
\usepackage{amssymb}
\usepackage{esint} 
\usepackage{commath} 
\usepackage{units} 

\usepackage{array}

\newcolumntype{P}[1]{>{\centering\arraybackslash}p{#1}}
\newcolumntype{M}[1]{>{\centering\arraybackslash}m{#1}}

\usepackage[caption=false,font=footnotesize]{subfig}
\usepackage{multirow}

\usepackage[unicode=true,pdfusetitle,%
 pdfauthor={Gustav Eje Henter, Jaime Lorenzo-Trueba, Xin Wang, Junichi Yamagishi},%
 pdfkeywords={Controllable speech synthesis, latent variable models, autoencoders, variational inference,VQ-VAE},%
 bookmarks=true,bookmarksnumbered=true,bookmarksopen=false,%
 breaklinks=true,pdfborder={0 0 0},backref=false,colorlinks=true,citecolor=blue]{hyperref}

\newcommand{\bb}[1]{\boldsymbol{#1}}
\newcommand{\seq}[1]{\underline{#1}}
\newcommand{\est}[1]{\widehat{#1}}
\newcommand{\given}[2]{\left(#1\,\middle|\,#2\right)}
\newcommand{\givenpar}[3]{\left(#1\,\middle|\,#2;\,#3\right)}

\newcommand{\real}{\mathbb{R}}

\newcommand{\cL}{\mathcal{L}}

\newcommand{\data}{\mathcal{D}}

\DeclareMathOperator*{\argmin}{argmin}
\DeclareMathOperator*{\argmax}{argmax}

\newcommand{\kldiv}[2]{D_{\mathrm{KL}}\left(#1\,\middle|\middle|\,#2\right)}
\newcommand{\avg}[2][]{\mathbb{E}_{#1}\left[#2\right]}

\newcommand{\esttheta}{\est{\boldsymbol{\theta}}}
\newcommand{\estmu}{\est{\boldsymbol{\mu}}}
\newcommand{\estz}{\est{\boldsymbol{z}}}
\newcommand{\estphi}{\est{\boldsymbol{\varphi}}}
\newcommand{\estpsi}{\est{\boldsymbol{\psi}}}
\newcommand{\estZ}{\est{\mathcal{Z}}}

\newcommand{\Lseq}{\seq{\bb{L}}}
\newcommand{\Xseq}{\seq{\bb{X}}}
\newcommand{\lseq}{\seq{\bb{l}}}
\newcommand{\xseq}{\seq{\bb{x}}}

\newcommand{\elbo}{\underline{\mathcal{L}}}

\let\oldmarginpar\marginpar
\renewcommand\marginpar[1]{\-\oldmarginpar[\raggedleft\footnotesize #1]%
{\raggedright\footnotesize #1}}

\begin{document}
%
\title{Deep Encoder-Decoder Models for Unsupervised Learning of Controllable Speech Synthesis}
%
%
%

\author{Gustav~Eje~Henter,~\IEEEmembership{Member,~IEEE,}
        Jaime~Lorenzo-Trueba\textsuperscript{\textdaggerdbl},~\IEEEmembership{Member,~IEEE,}
        Xin~Wang,~\IEEEmembership{Student Member,~IEEE,}
        and~Junichi~Yamagishi,~\IEEEmembership{Senior Member,~IEEE}
\thanks{Manuscript last revised July 30, 2018.}
\thanks{This research was carried out while all authors were with the Digital Content and Media Sciences Research Division at the National Institute of Informatics, 2-1-2 Hitotsubashi, Chiyoda-ku, Tokyo 101-8430, Japan.}
\thanks{G.~E.~Henter is with the Department of Speech, Music and Hearing (TMH) at KTH Royal Institute of Technology, 100 44 Stockholm, Sweden.\ (e-mail: \href{mailto:ghe@kth.se}{ghe@kth.se})}
\thanks{J.~Lorenzo-Trueba is with Amazon.com in Cambridge, U.K.\ (e-mail: \href{mailto:jaime@nii.ac.jp}{jaime@nii.ac.jp})}
\thanks{X.~Wang is with the Digital Content and Media Sciences Research Division at the National Institute of Informatics, Japan.\ (e-mail: \href{mailto:wangxin@nii.ac.jp}{wangxin@nii.ac.jp})}
\thanks{J.~Yamagishi is with the Digital Content and Media Sciences Research Division at the National Institute of Informatics, Japan, as well as with the Centre for Speech Technology Research at the University of Edinburgh, 10 Crichton Street, Edinburgh EH8 9AB, U.K.\ (e-mail: \href{mailto:jyamagis@nii.ac.jp}{jyamagis@nii.ac.jp})}
\thanks{\textsuperscript{\textdaggerdbl} Work performed prior to joining Amazon.}}

%
%

\markboth{Preprint. Work in progress.}{Preprint. Work in progress.}

%


\maketitle

\begin{abstract}
Generating versatile and appropriate synthetic speech requires control over the output expression separate from the spoken text. Important non-textual speech variation is seldom annotated, in which case output control must be learned in an unsupervised fashion. In this paper, we perform an in-depth study of methods for unsupervised learning of control in statistical speech synthesis. For example, we show that popular unsupervised training heuristics can be interpreted as variational inference in certain autoencoder models. We additionally connect these models to VQ-VAEs, another, recently-proposed class of deep variational autoencoders, which we show can be derived from a very similar mathematical argument. The implications of these new probabilistic interpretations are discussed. We illustrate the utility of the various approaches with an application to acoustic modelling for emotional speech synthesis, where the unsupervised methods for learning expression control (without access to emotional labels) are found to give results that in many aspects match or surpass the previous best supervised approach.
\end{abstract}


%
\IEEEpeerreviewmaketitle

%
%
%

\section{Introduction}
\label{sec:introduction}
\IEEEPARstart{T}{ext} to speech (TTS) is the task of turning a given text into an audio waveform of the text message being spoken out loud.
While speech waveforms have a very high bitrate (e.g., 705,600 bits per second for CD-quality audio), the spoken text only accounts for a handful of these bits, perhaps 50 or 100 bits per second \cite{vankuyk2017information}. A major challenge of text-to-speech synthesis is thus to fill in the additional bits in the audio signal in an appropriate and convincing manner. This is not an easy task, as speech features have complex interdependencies \cite{henter2014measuring}. Furthermore, much of the excess acoustic variation in speech is not completely random and incidental, but conveys additional side-information of relevance to communication. The acoustics may, for instance, reflect characteristics such as speaker identity, speaker condition, speaker mood and emotion, pragmatics (via emphasis and intonation), the acoustic environment, and properties of the communication channel (microphone characteristics, room acoustics). Neither of these are determined by the spoken text.

Ideally, the acoustic cues and variability encountered in natural speech should not only be replicated in the acoustics to make the synthesis more convincing, but also be adjustable to create flexible and expressive synthesisers, and ultimately enhance communication between man and machine.
Unfortunately, this is not the case today.
Most statistical parametric speech synthesis approaches are based on supervised learning, and only account for the variation that can be directly explained by the annotation provided. Any deviations from the conditional mean as predicted from annotated labels is assumed to be random and largely uncorrelated, regardless of any structure or information it may possess.

At synthesis time, recreating the lost variability by drawing random samples from fitted Gaussian models has been found to be a poor strategy from a perceptual point of view, cf.\ \cite{uria2015modelling}, wherefore the predicted average speech features are used in synthesis instead; in fact, acoustic models must be highly accurate before random sampling outperforms the average speech \cite{henter2014measuring}. Using the model mean for synthesis makes the same utterance sound exactly identical every time it is synthesised (unlike when humans speak), and is still likely to give rise to artefacts, for instance widened formant bandwidths when using spectral or cepstral acoustic feature representations.

In theory, salient variation beyond the text could be annotated in the database, enabling the acoustic effects of the additional labels to be learned during training and controlled during synthesis. However, speech annotation is laborious, difficult, and often subjective. This makes it costly to obtain sufficient amounts of data where non-text variation has been annotated accurately. Instead, synthesis practise has focussed on reducing the amount of (unhandled) acoustic variability by recording TTS databases of single talkers reading text in a consistent neutral tone. The use of such data for building synthesisers may benefit segmental acoustic quality, but likely contributes to the flat and detached delivery that many text-to-speech systems suffer from. Several publications \cite{fan2015multi,vandenoord2016wavenet,li2016multi,luong2017adapting} have meanwhile highlighted the potential benefits of acoustic variation (at least when annotated), for instance \cite{luong2017adapting} presenting multi-speaker synthesisers that are more accurate than could be expected from training on any single speaker in the database alone and additionally allow control over properties of the generated speech, such as the speaker's voice.

This paper considers a number of alternatives to the standard approach outlined above. The common theme is to investigate and connect methods that attempt to explicitly account for the effects of \emph{unannotated} variation in the data. These methods are able to learn synthesisers with controllable output acoustics (beyond the effects of the input text), albeit without an a-priori labelling of the perceptual effects of the learned control; this can be seen as an important, though not sufficient, step to eventually enable flexible speaking systems that respond appropriately to communicative context. 
Mathematically, our perspective is that of probabilistic modelling, specifically the theory of latent variables, and a major part of the work is to establish theoretical connections between practical approaches and principles of statistical estimation. Our main scientific contributions can be summarised as follows:
\begin{enumerate}
\item We use variational methods to show that several prior methods for learning controllable models from data with unannotated variation -- the training heuristic used in \cite{abdel2013fast,xue2014fast,watts2015sentence,luong2017adapting,arik2017deep,taigman2018voiceloop}, as well as so-called VQ-VAEs from \cite{van2017neural} -- can be interpreted as approximate maximum-likelihood approaches, and elucidate the approximations involved.
\item We introduce and detail various theoretical connections between the techniques in \cite{watts2015sentence,luong2017adapting,arik2017deep,taigman2018voiceloop} and encoder-decoder models, particularly VQ-VAEs.
\item We consider ways in which prior information can be integrated into the heuristic approaches (which lack an explicit prior distribution).
\item We use a large database of emotional speech to perform objective and subjective empirical evaluations of the heuristic approaches (with and without prior information) against comparable VQ-VAEs and a competitive supervised system on the task of acoustic modelling. The unsupervised methods are found to produce equal or better results than the supervised approach.
\end{enumerate}
These contributions all extend preliminary work performed in \cite{henter2017principles}.

The remainder of this article is laid out as follows: Sec.\ \ref{sec:background} outlines relevant prior work while Sec.\ \ref{sec:theory} describes mathematical foundations. Sec.\ \ref{sec:insights} then presents novel interpretations of and connections between different encoder-decoder approaches. Sec.\ \ref{sec:experiments} recounts empirical evaluations performed on a database of emotional speech, while Sec.\ \ref{sec:conclusion} concludes.

\section{Prior Work}
\label{sec:background}
In this section, we introduce controllable speech synthesis (Sec.\ \ref{ssec:control}) and a wide variety of previous work of relevance to our contributions. We especially consider unsupervised learning of control (Sec.\ \ref{ssec:unsupervised}) and variational autoencoders (Sec.\ \ref{ssec:vaes}) and their use in speech generation (Sec.\ \ref{ssec:vaesintts}). We also give an introduction to prior work on emotional speech synthesis (Sec.\ \ref{ssec:emotionaltts}), as this is the control task considered in our experiments.

\subsection{Controllable Speech Synthesis}
\label{ssec:control}
All text-to-speech systems are in a sense controllable, since the input text influences the output audio. (Voice conversion, similarly, represents a speech synthesiser driven by speech rather than text.) 
By \emph{controllable speech synthesis}, however, we refer to speech synthesisers that enable additional output control beyond the words alone, such that the same text can be made to be spoken in several, perceptually distinct ways.

Early, rule-based parametric speech synthesisers typically exposed many control knobs (``speech parameters'') relating to speech articulation and pronunciation; the text-to-speech aspect was simply a set of rules for how these knobs were to be moved in response to phonemes extracted from text \cite{klatt1987review}, and the resulting parameter trajectories could be manually edited in order to alter pronunciation.
Unit selection TTS can achieve control of any properties annotated in the database by including a term in the target cost to preferentially select units with labels similar to the user-selected control input. However, success depends heavily on the database having adequate coverage of the desired control configuration.

With the transition to statistical parametric speech synthesis (SPSS), \cite{zen2009statistical,king2011introduction} it became straightforward to learn to control synthesiser output, i.e., to learn a mapping from control inputs to acoustic outputs. This avoids having to design the signal generator to expose the desired speech properties to be controlled or manually tuning weight factors in the target cost, and typically achieves meaningful control from smaller training databases than unit-selection approaches.
The decision trees used in early SPSS systems can relatively easily incorporate additional categorical labels as phone- or frame-level inputs. Continuous-valued inputs can be quantised for decision-tree learning, and the quantisation threshold can be learned as well (e.g., through C4.5 \cite{quinlan1996improved}). So-called multiple regression HMMs (MR-HMMs) \cite{fujinaga2001multiple} were developed as a more refined method for continuous control of synthesiser output, by endowing each decision-tree node with a linear regression model that maps control inputs to acoustics. MR-HMMs and their extensions have been used for smoothly controlling properties such as speaking style \cite{masuko2004style,nose2007style} or articulation \cite{ling2013articulatory}.


\subsection{Learning Control Without Annotation}
\label{ssec:unsupervised}
The approaches covered in Sec.\ \ref{ssec:control} all rely on control either being manually designed, or learned in a supervised manner from annotated data. This paper, in contrast, considers the more difficult situation where salient speech variability has not been annotated, but we nonetheless wish to learn to account for and replicate such variability by adjusting some synthesiser control inputs separate from the input text.



Many approaches to this problem exist. Unlike, e.g., Jauk \cite{jauk2017unsupervised}, where the control space is defined by clustering training utterances based on pre-defined acoustic features, we concentrate on approaches that treat the unknown values of the hypothesised control parameters as if they were part of the set of unknown model parameters, and estimate all these unknowns through optimisation over the training data.
This will learn a synthesiser that allows the control over the most (mathematically) salient extra-linguistic speech variation, but provides no a-priori indication what perceptual aspects that will be controllable (or how).
One example of this approach is so-called cluster-adaptive training (CAT), introduced for automatic speech recognition (ASR) in \cite{gales2000cluster}. It can be seen as an extension of MR-HMMs to learning and optimising both decision-tree node regression models and their inputs. CAT has for instance been applied to learn expressive TTS with decision trees \cite{chen2012exploring}. However, the method does not include a joint optimisation over the regression tree structure, and the possible uncertainty in the determination of the control input values from the acoustics is ignored.

With modern synthesis techniques based on deep learning there have been multiple independent proposals to improve modelling by using backpropagation to jointly optimise the entire regression model (the unknown weights of one or more neural networks) together with its control inputs.
The idea was introduced for speaker adaptation in neural network ASR in \cite{abdel2013fast,xue2014fast} under the name ``discriminant condition codes'' (DCC), and was independently adapted for multi-speaker speech synthesis several times: first by Luong et al.\ \cite{luong2017adapting} and more recently by Ar{\i}k et al.\ \cite{arik2017deep} (Deep Voice 2) and Taigman et al.\ \cite{taigman2018voiceloop} (VoiceLoop).
In all cases, the result is that training and test speakers all are embedded in a low-dimensional speaker space. 
Independent of \cite{xue2014fast}, Watts et al.\ \cite{watts2015sentence} also proposed a mathematically identical setup and applied it to train a TTS acoustic model on a database of expressive speech, specifically children's audiobooks from \cite{king2016blizzard}. (The equivalence between \cite{xue2014fast} and \cite{watts2015sentence} was first pointed out in \cite{henter2017principles}.) Watts et al.\ learned a fixed input vector for each utterance in the data, calling the approach ``learned sentence-level control vectors''. Adjusting the control parameter input when synthesising from the trained system was found to adjust vocal effort (pitch and energy) in a nonlinear and non-uniform manner.

Sawada et al.\ \cite{sawada2017nitech} considered similar data but took a somewhat different approach, wherein a unique ``phrase code'' was assigned to each phrase in the training data through random draws from a high-dimensional Gaussian distribution; this code was then used as an input to the synthesiser alongside the features extracted from the text. For test sentences, the phrase code of the training-data phrase with the greatest similarity (as computed through by doc2vec \cite{le2014distributed}) to the text phrase to be spoken was used as the control parameters. (They also assigned ``word codes'' to each word in a similar manner.) This overall approach is similar to the approaches with learned input codes -- especially \cite{watts2015sentence} -- in that training-data segments were embedded in a fixed-dimensional space used to control the output, but here the embeddings were random rather than learned, and codes were predicted based on text rather than acoustics. Trained on children's audiobooks the resulting synthesiser achieved notably successful expression control and was one of the best-rated systems in the 2017 Blizzard Challenge \cite{king2017blizzard,sawada2017nitech}.

Luong et al.\ \cite{luong2017adapting} evaluated both random and learned input codes with different dimensionalities for representing speaker variation, and compared them to simple one-hot vector speaker codes. They found no major differences in subjective performance between the methods, though all were better than no adaptation. However, we note that this and other speaker-adaptation evaluations typically involve some degree of supervision, since it generally is pre-specified which utterances that came from each speaker.

In the last year, there have been efforts to learn unsupervised control in the (mostly) end-to-end Tacotron \cite{wang2017tacotron} TTS framework. Parallel to this paper being written, these demonstrated the use of encoders and decoders for prosody transfer across speakers (given similar text prompts) \cite{skerry2018towards} and more general style control \cite{wang2018style}. This extends and improves on preliminary work presented by the same group in \cite{wang2017uncovering}, which learned framewise rather than utterance-level control. Among other things, they demonstrate that the style-token approach in \cite{wang2018style} is capable of synthesis with high subjective quality even from 95\% noisy training data. They also demonstrated the use of a separately-learned speaker verification system as an encoder for controlling and adapting speaker identity \cite{jia2018transfer}.

\subsection{Variational Autoencoders}
\label{ssec:vaes}
Interestingly, all of the above proposals for unsupervised learning of controllable speech synthesis gloss over the issue that the actual values of any control inputs cannot be determined to exact certainty, since they are neither annotated nor observed. To properly account for the uncertainty regarding the unknown control inputs calls for the use of \emph{latent} (or \emph{hidden}) \emph{variables} associated with each datum. The fundamental idea is simply to model the unknown quantities and their uncertainty as random variables.
We can then use the theory of probability and estimation to make inferences about these unobserved variables.
In practice, the mathematics are very similar to Bayesian probability, but the prior and posterior distributions pertain to (local) control inputs, not to the (global) model parameters, which may still be treated in a frequentist manner.


Latent-variables are ubiquitous in speech modelling, with two examples being the component in a mixture model and the unobservable state variable in hidden Markov models (HMMs) \cite{bishop2006pattern,rabiner1989tutorial}.
Training algorithms for these latent-variable approaches are usually derived from the expectation-maximisation (EM) framework \cite{dempster1977maximum}.
However, the expressiveness of these classical methods is often quite limited, and new setups generally require careful, manual derivation of update equations, which often is prohibitively difficult for more complex and interesting models.

A recent idea is to harness the power of deep learning to describe and train more flexible latent-variable models. 
Using techniques similar to \cite{dempster1977maximum}, Henter et al.\ \cite{henter2017principles} showed that, for the special case of EM-like alternate optimisation, the heuristic methods \cite{abdel2013fast,xue2014fast,watts2015sentence,luong2017adapting,arik2017deep,taigman2018voiceloop} can be seen as ``poor man's latent variables'' that can learn a complex mapping from latent to observable variables but ignore any uncertainty in the latent space.
A more full-fledged example of deep learning of latent variables is so-called \emph{variational autoencoders} (VAEs) \cite{kingma2014auto,rezende2014stochastic}. They use neural networks to parameterise both how observations depend on continuous latent variables (control inputs) along with the act of inferring latent-variable distributions from observations. VAEs are considered autoencoders since the inference process can be seen as encoding an observation into a latent variable value (or distribution) while the generation can be seen as decoding the latent variable back to the observation domain. We elaborate on this connection in Sec.\ \ref{ssec:vaemaths}. Furthermore, the two mappings can be learned tractably and jointly through gradient descent \cite{doersch2016tutorial}, in contrast to some mathematically similar models such as Helmholtz machines \cite{dayan1995helmholtz}.


A practical issue with VAEs is that they sometimes fail to learn to make proper use of the latent variables to explain the observed variation: in that case, the estimated control inputs do not change appreciably over the training data (their inferred distributions are highly overlapping) and exert little influence over model outputs, cf.\ \cite{blaauw2016modeling}. Chen et al.\ \cite{chen2016variational}, Husz\'{a}r \cite{huszar2017maximum}, and Graves et al.\ \cite{graves2018associative} provide lucid discussions of this problem. This has been called ``posterior collapse'' in \cite{van2017neural}, although it does \emph{not} mean that the posterior collapses to a point -- just that the posterior collapses to the same distribution (which is also the prior) regardless of the observation made. 
A recent proposal to combat this issue is to quantise the encoder output through a vector-quantisation (VQ) step, such that the inferred value of the hidden variable for an observation is taken from a finite codebook. The resulting construction is called \emph{VQ-VAE}, and was introduced in \cite{van2017neural}.
While the regular VAEs objective function penalises the variational posterior diverging from the prior (which can force ``posterior collapse''), this penalty reduces to a constant for the VQ-VAE, and thus does not affect learning.
Although the fact that only a single codebook vector is used for each observation means that any uncertainty in the inference step is not represented explicitly, we show in Sec.\ \ref{ssec:vqvaes} that the mathematics still can be derived from the same latent-variable principles that underpin regular VAEs. VQ-VAEs might use discrete latent variables, but these latents are nonetheless embedded in a continuous Euclidean space.


While Gaussian mixture models and HMMs also consider discrete latent variables that are in some sense embedded (through their mean vectors) in a vector space, VQ-VAEs let the latent vectors occupy a space different from that of the observations. The VQ-VAE mapping from latent space to observation space is furthermore strongly nonlinear, which differentiates it from constructions like subspace GMMs \cite{povey2010subspace}.

Variational autoencoders also resemble recently-popular generative adversarial networks (GANs) \cite{goodfellow2014generative}, in that the latter also use a random latent variable to explain variation in the observations through a highly-nonlinear mapping parameterised by a neural network. However, VAEs map latent variable values to output distribution parameters, whereas GANs map latent samples directly to observations. Parameter estimation in GANs is also more challenging, since one seeks a Nash equilibrium of a game between two agents, rather than an optimum of a fixed objective function as in VAEs. A taxonomy of different generative models such as VAEs and GANs, along with connections between them, is provided in \cite{goodfellow2016nips}.
In Sec.\ \ref{sec:insights} this paper, we bring the widely-used heuristic from Sec.\ \ref{ssec:unsupervised} (DCC/sentence-level control vectors) into the fold, by describing its connections to VAEs and latent-variable models.

\subsection{Variational Autoencoders in Synthesis}
\label{ssec:vaesintts}
Variational autoencoders have seen a number applications to speech generation. For example, \cite{blaauw2016modeling,hsu2016voice,hsu2017voice,kameoka2018acvaevc} all consider applying VAEs to each frame in an acoustic analysis of speech, with the intention of learning to encode something similar to phonetic identity in the absence of transcription. In \cite{hsu2016voice,hsu2017voice}, this was used to identify matching data frames for non-parallel voice conversion.
\cite{hsu2017learning,hsu2017unsupervised} used VAEs to separate and manipulate both speaker and phone identities, though without generating or evaluating speech audio. Very recently \cite{akuzawa2018expressive} used VAEs to identify sentence-level latent variables in the VoiceLoop \cite{taigman2018voiceloop} framework.

VAEs have also been applied to speech waveform modelling, typically based on generalisations of basic VAEs to sequence models such as \cite{fabius2014variational,chung2015recurrent,fraccaro2016sequential,marino2018general}.
While \cite{chung2015recurrent,fraccaro2016sequential,marino2018general} all contain applications to speech data, only Chung et al.\ \cite{chung2015recurrent} considered speech signal generation. Unfortunately, the perceptual quality of random waveforms sampled from their model is poor: there is a lot of static, and no intelligible speech is produced, since the models are not conditioned on an input text. Much better segmental quality has been demonstrated by generating signals using WaveNet \cite{vandenoord2016wavenet}. In a standard WaveNet the next-step distribution only depends on the previous waveform in the receptive field and possible conditioning information, with no hidden state. Other successful neural networks for waveform generation include SampleRNN \cite{mehri2016samplernn} and WaveRNN \cite{kalchbrenner2018efficient}, which contain a deterministic (hidden) RNN state. The VQ-VAE paper \cite{van2017neural} combines these breakthroughs (specifically WaveNet) with VAEs, using strided convolutions to downsample and encode raw audio into discrete quantisation indices with a WaveNet-like architecture for decoding. This approach was able to reproduce high-quality versions of encoded waveforms, and the quantisation indices were additionally found to be closely related to phones, providing a compelling demonstration of unsupervised acoustic unit discovery.

Wang \cite[Ch.\ 7]{wang2018fundamental} investigated VQ-VAEs for F0 modelling on the utterance, mora, and phone levels in Japanese TTS, coupled with a \emph{linguistic linker} to predict VQ-VAE codebook indices from linguistic features. It was found that a combined VQ-VAE approach on the mora and phone levels performed objectively and subjectively on par with a larger deep, autoregressive F0 model \cite{wang2018autoregressive} without explicit latent variables.

Different from the prior work above, but similar to the heuristics \cite{abdel2013fast,xue2014fast,watts2015sentence,luong2017adapting,arik2017deep,taigman2018voiceloop} in Sec.\ \ref{ssec:unsupervised}, this paper considers \mbox{(VQ-)VAE} approaches that model and encode utterance-wide, non-phonetic information that complements the known transcription.

The work on speech synthesis with global style tokens (GSTs) in \cite{wang2018style} has many similarities to VQ-VAEs and encoder-decoder based synthesis. While the global style tokens are initialised as random vectors (like in, e.g.,\cite{sawada2017nitech}), only a limited, fixed number of style tokens is used, reminiscent of a vector-quantiser codebook. Unlike VQ-VAEs, however, the style-token approach uses attention to obtain a set of positive interpolation weights between the different tokens. This means that utterances in practice can fall on a continuum in token space, similar to the heuristic approaches in Sec.\ \ref{ssec:unsupervised}.
Another difference is that the encoders in \cite{skerry2018towards,wang2018style,jia2018transfer} do not have access to the text features, in contrast to the heuristic and VQ-VAE approaches studied in this paper, which make use of both acoustic and text-derived features in encoding.

\subsection{Emotional Speech Synthesis}
\label{ssec:emotionaltts}
The experiments in this paper consider speech synthesis from a large corpus of acted emotional speech, described in \cite{lorenzo2018investigating}.
The importance of emotional expression in speech synthesis can be seen in, e.g., the 2016 Blizzard Challenge 
\cite{king2016blizzard}, where suitably accounting for the expressive nature of the data was a common element of the most successful entries.


There have been successful demonstrations of emotional speech synthesis with speech generation based on unit selection (including hybrid speech synthesis) \cite{barra2010analysis,erro2010emotion,tsiakoulis2016affective} as well as through SPSS with decision trees \cite{yamagishi2005acoustic,nose2013intuitive,lorenzo2015emotion,cabral2016adapt,do2016hybrid}.
Most of these consider a relatively limited number of discrete emotional classes, from binary (e.g., neutral vs.\ affective as in \cite{tsiakoulis2016affective}) to the ``big six'' (anger, disgust, fear, happiness, sadness and surprise, as considered in \cite{barra2010analysis,erro2010emotion,cabral2016adapt}); \cite{nose2013intuitive}, which investigates continuous emotional-intensity control with MR-HMMs, is an exception.
Applications of methods based on neural-networks to emotional speech synthesis are less common, though there are a few examples \cite{henter2017principles,lorenzo2018investigating} from the last year. This article builds on these two publications and considers the same data in the experiments.

\section{Mathematical Background}
\label{sec:theory}
This section introduces the mathematical preliminaries of speech synthesis as necessary for the novel insights described in Sec.\ \ref{sec:insights}. In particular, Sec.\ \ref{ssec:estimation} outlines controllable speech synthesis through latent variables, while remaining sections describe the fundamental theory of variational inference (Sec.\ \ref{ssec:vi}) and variational autoencoders in general (Sec.\ \ref{ssec:vaemaths}).

\subsection{Controlling Speech Synthesis Through Latent Variables}
\label{ssec:estimation}
Mathematically, statistical parametric speech synthesis is usually formulated as a regression problem.
The central statistical modelling task is to map an input sequence $\lseq$ of text-based (``linguistic'') features to a sequence $\xseq$ of acoustic features (``speech parameters'') that control a waveform generator (vocoder).%
\footnote{In this text, bold symbols signify vectors or matrices; the underline denotes a time sequence $\lseq=\left(\bb{l}_1,\,\ldots,\,\bb{l}_T\right)$. Capital letters identify random variables, while corresponding lowercase quantities represent specific, non-random outcomes of those variables.}
Since human speech is stochastic even for a given text and control input (cf.\ \cite{henter2014measuring}), we typically want to map the input $\lseq$ to an entire distribution $\Xseq(\lseq)$ of acoustic feature sequences $\xseq$.
This mapping is learned from a parallel corpus of text and speech using statistical methods.
The linguistic features $\lseq$ in the mapping are extracted deterministically from input text by a (typically language-dependent) so-called front-end. While the front-end traditionally has been designed rather than learned, this is starting to change, with a number of frameworks \cite{wang2017tacotron,sotelo2017char2wav,ping2018deep,taigman2018voiceloop} learning to predict acoustics directly from sequences of characters or phones. Similarly, the waveform generator is traditionally a fixed, designed component, for example STRAIGHT \cite{kawahara2006straight} or WORLD \cite{morise2016world}, to whose control interface the acoustic feature representation is tied. However, learned (neural) vocoders have recently achieved impressive results, e.g., \cite{shen2018natural}.
Thus, while it is possible to learn both the front-ends and vocoders, only the central linguistic-to-acoustic mapping is consistently learned from speech data.%
\footnote{For all the interest in waveform-level speech synthesis, it is worth noting that \cite{shen2018natural} -- the current state of the art in text-to-speech signal quality -- still solves a statistical parametric speech synthesis problem. The difference in speech quality comes from matched training of a learned vocoder instead of synthesising waveforms with the Griffin-Lim algorithm as in \cite{wang2017tacotron}.}

Let $\data=\left\{\lseq^{(n)},\,\xseq^{(n)}\right\}_{n=1}^{N}$ be a dataset of $N$ aligned linguistic (input) and acoustic (output) data sequences, which are assumed to be independent and identically distributed draws from a joint distribution of $\Lseq$ and $\Xseq$. Let further $f_{\Xseq\mid{}\Lseq}\givenpar{\xseq}{\lseq}{\bb{\theta}}$ be a parametric model describing the probability of output $\Xseq$ given $\Lseq$. To estimate the unknown model parameters $\bb{\theta}$ it is standard to use maximum-likelihood estimation
\begin{align}
\esttheta_{\mathrm{ML}}(\data)
& = \argmax_{\bb{\theta}} \cL\given{\bb{\theta}}{\data} \\
\cL\given{\bb{\theta}}{\data}
& = \sum_{n=1}^{N} \ln f_{\Xseq\mid{}\Lseq}\givenpar{\xseq^{(n)}}{\lseq^{(n)}}{\bb{\theta}}
\label{eq:loglike}
\text{.}
\end{align}

To achieve control over how the text message encoded by $\lseq$ is spoken, we add a second input representing control parameters, $\bb{z}$. While one could envision using a sequence $\seq{\bb{z}}\in\real^D$ of control inputs that may change throughout an utterance, we only develop the mathematics for the case when this input is constant for each data sequence, and thus can be represented by a single vector $\bb{z}$. If this control signal has been annotated as $\bb{z}^{(n)}$ for each training data sequence it is straightforward to train a controllable synthesiser by maximising the conditional likelihood
\begin{align}
\cL\given{\bb{\theta}}{\data}
& = \sum_{n=1}^{N} \ln f_{\Xseq\mid{}\Lseq,\,\bb{Z}}\givenpar{\xseq^{(n)}}{\lseq^{(n)},\,\bb{z}^{(n)}}{\bb{\theta}}
\text{.}
\end{align}
Changing the control signal will then cause the output distribution to be more similar to the examples with similar annotated control-input values, assuming learning was successful.

The situation becomes more interesting if the control parameter is a latent (unobserved) variable. A general and principled approach is to treat the unknown control input as a random variable $\bb{Z}$ which is jointly distributed with $\Xseq$ as in 
\begin{align}
f_{\Xseq,\,\bb{Z}\mid{}\Lseq}\givenpar{\xseq,\,\bb{z}}{\lseq}{\bb{\theta}}
& = f_{\Xseq\mid{}\Lseq,\,\bb{Z}}\givenpar{\xseq}{\lseq,\,\bb{z}}{\bb{\theta}}
f_{\bb{Z}\mid{}\Lseq}\givenpar{\bb{z}}{\lseq}{\bb{\theta}}
\text{,}
\end{align}
where $f_{\bb{Z}\mid{}\Lseq}$ is a conditional prior for $\bb{Z}$. To perform maximum-likelihood parameter estimation in the presence of this latent variation one marginalises out the unknown random variable, and thus maximises 
\begin{align}
\cL\given{\bb{\theta}}{\data}
& = \sum_{n=1}^{N} \ln \int f_{\Xseq,\,\bb{Z}\mid{}\Lseq}\givenpar{\xseq^{(n)},\,\bb{z}}{\lseq^{(n)}}{\bb{\theta}} \dif{}\bb{z}
\label{eq:marglike}
\text{;}
\end{align}
this is termed the \emph{marginal likelihood} or the \emph{model evidence}, but is merely another way of writing $f_{\Xseq\mid{}\Lseq}$ from Eq.\ \eqref{eq:loglike}.

To generate speech from a latent-variable model like this, there are two conceivable $\Xseq$-distributions to consider. One could use the same marginalisation principle as in Eq.\ \eqref{eq:marglike} and generate speech based on $f_{\Xseq\mid{}\Lseq}$ (i.e., after integrating out $\bb{Z}$). However, the integral is frequently intractable, as discussed in the next paragraph. Moreover, this does not allow control of the output speech $\xseq$. For these reasons we exclusively consider output generation from the $\Xseq$-distribution conditioned on $\bb{Z}$, $f_{\Xseq\mid{}\Lseq,\,\bb{Z}}$. By adjusting the input $\bb{z}$-value, the same text may then be spoken in (statistically) distinct ways.

\subsection{Variational Inference}
\label{ssec:vi}
Unfortunately, the integral in Eq.\ \eqref{eq:marglike} is only tractable to evaluate for quite basic models, which tend to be too simplistic to allow an acceptable description of reality. To fit more advanced statistical models, approximations must be made. Some approximation techniques rely on numerical methods for estimating the value of the integral, e.g., through Monte-Carlo sampling. In this paper, however, we consider analytical approximations based on variational principles, where a parametric and tractable approximation $q(\bb{z};\,\bb{\varphi})$ is used in place of the intractable true posterior $f_{\bb{Z}\mid{}\Xseq,\,\Lseq}$. Instead of maximising the likelihood $\cL$ directly, one then maximises a lower bound $\elbo$ on it, sometimes called the \emph{evidence lower bound} (ELBO). Specifically, one can show \cite[Sec.\ 10.1]{bishop2006pattern} that 
\begin{align}
\ln f_{\Xseq\mid{}\Lseq}\givenpar{\xseq}{\lseq}{\bb{\theta}}
& = \kldiv{q}{f_{\bb{Z}\mid{}\Xseq,\,\Lseq}}
+ \elbo\given{\bb{\theta},\,\bb{\varphi}}{\xseq,\,\lseq}
\label{eq:variational}
\text{,}
\end{align}
where
\begin{align}
\kldiv{q}{f_{\bb{Z}\mid{}\Xseq,\,\Lseq}}
& = \int q(\bb{z};\,\bb{\varphi}) \ln\frac{q(\bb{z};\,\bb{\varphi})}{f_{\bb{Z}\mid{}\Xseq,\,\Lseq}\givenpar{\bb{z}}{\xseq,\,\lseq}{\bb{\theta}}} \dif{}\bb{z}
\label{eq:kld}
\end{align}
is the Kullback-Leibler divergence (or KLD) and
\begin{align}
\elbo\given{\bb{\theta},\,\bb{\varphi}}{\xseq,\,\lseq}
& = \int q\left(\bb{z};\,\bb{\varphi}\right) \ln \frac{f_{\Xseq,\,\bb{Z}\mid{}\Lseq}\givenpar{\xseq,\,\bb{z}}{\lseq}{\bb{\theta}}}{q\left(\bb{z};\,\bb{\varphi}\right)} \dif{}\bb{z}
\label{eq:elbo}
\end{align}
is the evidence lower bound. Since the KLD between two distributions satisfies $\kldiv{p}{q}\geq0$, with equality if and only if $p=q$, the desired bound $\cL\geq\elbo$ follows. This bound can be applied to every term in Eq.\ \eqref{eq:loglike} with a separate $q$-distribution $q(\bb{z};\,\bb{\varphi}^{(n)})$ for each datapoint to lower-bound the entire training-data likelihood.

If $q$ is chosen cleverly, the integral in Eq.\ \eqref{eq:elbo} can sometimes be evaluated analytically. One can then identify a parameter estimate $\esttheta_{\mathrm{VI}}$ and a set of per-datum $q$-distribution parameters $\bb{\varphi}^{\star(n)}$ (producing the variational posteriors $q^{\star(n)}$) that jointly maximise $\elbo$. This framework provides the basis for optimising and using powerful statistical models through the use of an approximate latent posterior. The difference between the optimal lower bound $\elbo$ and the optimal (log-)likelihood $\cL$ of the model without the variational approximation is given by $\kldiv{p}{q^\star}$ and is referred to as the \emph{approximation gap} \cite{cremer2018inference}.

\subsection{Variational Autoencoders}
\label{ssec:vaemaths}
The main idea of variational autoencoders \cite{kingma2014auto,rezende2014stochastic} is to use neural networks to parameterise not only the output-distribution dependence on latent-variable values, but also the act of latent-variable inference, and then learn these two networks simultaneously.
Like in variational inference in general, we approximate the true latent posterior $f_{\bb{Z}\mid{}\Xseq,\,\Lseq}$ by a variational posterior $q$, but instead of optimising the set $\left\{\bb{\varphi}^{(n)}\right\}$ to identify a different posterior distribution $q^{\star(n)}$ for each datapoint, these multiple optimisations are replaced by a single function $q_{\bb{Z}\mid{}\Xseq,\,\Lseq}\givenpar{\bb{z}}{\xseq,\,\lseq}{\bb{\varphi}}$ (here a neural network) that simply maps the values of $\xseq$ and $\lseq$ to (parameters of) an approximate posterior $q$.%
\footnote{Please note that $\bb{\varphi}$ now denotes a set of neural network weights that define a mapping from $\xseq$ and $\lseq$ to distribution parameters, rather than distribution parameters themselves as in Sec.\ \ref{ssec:vi}.}
This function $q_{\bb{Z}\mid{}\Xseq,\,\Lseq}$, parameterised by the network weights $\bb{\varphi}$, is sometimes called the \emph{inference network}, the \emph{recognition network}, or the \emph{encoder} and is distinct from the previously-introduced conditional output distribution $f_{\Xseq\mid{}\Lseq,\,\bb{Z}}$ (sometimes called the \emph{decoder}) that is parameterised by $\bb{\theta}$.


Given the parameterised inference $q_{\bb{Z}\mid{}\Xseq,\,\Lseq}$ defined above, one can show \cite{kingma2014auto,doersch2016tutorial} that
\begin{multline}
\ln f_{\Xseq}\left(\xseq;\,\bb{\theta}\right)
- \kldiv{q_{\bb{Z}\mid{}\Xseq}}{f_{\bb{Z}\mid{}\Xseq}} \\
= \avg[\bb{Z}\sim{}q_{\bb{Z}\mid{}\Xseq}]{ \ln f_{\Xseq\mid{}\bb{Z}}\givenpar{\xseq}{\bb{Z}}{\bb{\theta}} }
- \kldiv{q_{\bb{Z}\mid{}\Xseq}}{f_{\bb{Z}}}
\label{eq:vae}
\text{,}
\end{multline}
where we for succinctness have suppressed the dependence on $\lseq$. (Strictly speaking, our main consideration is \emph{conditional VAEs}, or C-VAEs, where every distribution additionally is conditioned on an input such as $\lseq$, but this difference is not of importance to the exposition.) The right-hand side in the equation is a lower bound on the likelihood (since the KLD on the left-hand side cannot be negative) which, it turns out, can be optimised efficiently using stochastic gradient ascent for certain choices of prior $f_{\bb{Z}\mid{}\Lseq}$ and approximate posterior $q$. A common choice \cite{kingma2014auto} is to take both distributions to be Gaussian; in this article we will additionally assume that the conditional output distribution $f_{\Xseq\mid{}\Lseq,\,\bb{Z}}$ is an isotropic Gaussian. 

The act of replacing individual optimisations by the regression problem of finding the weights $\bb{\varphi}$ in VAEs is sometimes called \emph{amortised inference}, since it amortises the computational cost of the separate optimisations (inferring $q^{(n)}$) over the entire training. (See \cite{shu2018amortized,cremer2018inference} for in-depth explanations.) Since the posterior parameters predicted by the learned $q$-function may not be optimal for each datapoint, VAEs will in practise usually not reach the same performance as the theoretically optimal $\elbo$ attained using $q^{\star(n)}$. The difference between the ELBO value attained by the VAE and the maximal ELBO possible under the chosen family of approximate posteriors $q$ is known as the \emph{amortisation gap} \cite{cremer2018inference}, and is added to the approximation gap due to the use of the approximate variational posterior defined in Sec.\ \ref{ssec:vi}.
\begin{figure}[!t]
\includegraphics[width=\columnwidth]{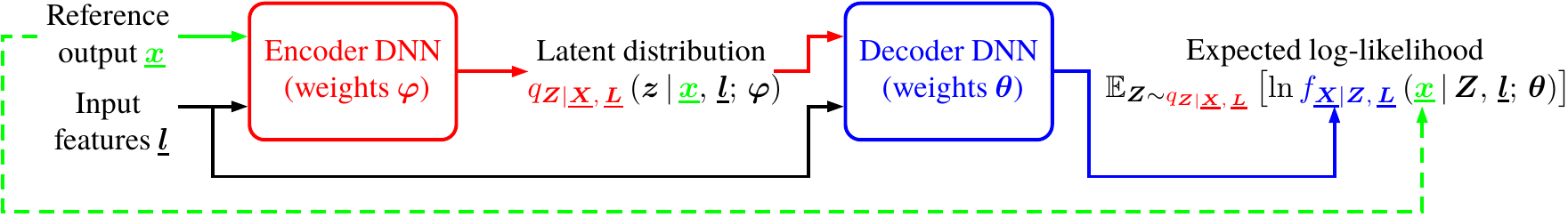}%
\caption{Conditional variational autoencoder training.}
\label{fig:vaeschematic}
\end{figure}


The ``autoencoder'' part of ``variational autoendcoders'' comes from the observation that $q_{\bb{Z}\mid{}\Xseq,\,\Lseq}\givenpar{\bb{z}}{\xseq,\,\lseq}{\bb{\varphi}}$ essentially encodes $\xseq$ into a latent variable $\bb{z}$, such that the original $\xseq$ is maximally likely to be recovered from (samples from) $q_{\bb{Z}\mid{}\Xseq,\,\Lseq}$, as seen in the expectation in Eq.\ \eqref{eq:vae}. This is illustrated conceptually in Fig.\ \ref{fig:vaeschematic}. Also note that the two terms on the right-hand side of Eq.\ \eqref{eq:vae} pull in different directions during maximisation: the first term is trying to make the approximate posterior $q_{\bb{Z}\mid{}\Xseq,\,\Lseq}$ resemble the true posterior as much as possible, while the second instead prioritises $q$ not straying too far from the given prior distribution. If our model class $f_{\Xseq\mid{}\Lseq,\,\bb{Z}}$ is sufficiently powerful to describe the observations well without depending on $\bb{z}$ as an input, the learned latent variables are likely to stay close to the prior and exert minimal influence on the observation distribution \cite{huszar2017maximum}. This is a common failure mode of VAEs, and is especially undesirable when learning output control.

To reduce the risk of not learning a useful latent-variable representation (``posterior collapse''), one can introduce a weight between the two terms in Eq.\ \eqref{eq:vae}, yielding so-called $\beta$-VAEs \cite{higgins2016beta}, which can also be annealed \cite{bowman2016generating}. This is straightforward to implement, but is not easy to motivate on probabilistic grounds and can not generally be interpreted as a lower bound on the marginal likelihood \cite{hoffman2017beta}. Alternatively, one might reduce the capacity/flexibility of the decoder model $f_{\Xseq\mid{}\bb{z},\,\lseq}$, for instance by modelling speech parameters with a simple Gaussian distribution as in the experiments in Sec.\ \ref{sec:experiments}.
VQ-VAEs were conceived as a third option for easily learning meaningful and informative latent representations.

\section{Theoretical Insights}
\label{sec:insights}
This section presents and discusses the main theoretical developments of this paper. In particular, Sec.\ \ref{ssec:vqvaes} describes a new probabilistic understanding of VQ-VAEs, Sec.\ \ref{ssec:heuristics} likewise introduces a variational derivation of the heuristic methods from \cite{abdel2013fast,xue2014fast,watts2015sentence,luong2017adapting,arik2017deep,taigman2018voiceloop} and connects these to other autoencoder models, while Sec.\ \ref{ssec:prior} discusses how prior information might be incorporated into the heuristic models. To the best of our knowledge, all of these contributions are new.

\subsection{A Variational Interpretation of VQ-VAEs}
\label{ssec:vqvaes}
VQ-VAEs were introduced in \cite{van2017neural} as a method of training VAEs when $\bb{Z}$ is a discrete random variable from a \emph{codebook} $\mathcal{Z}=(\bb{z}_1,\,\ldots,\,\bb{z}_M)$, a finite set of vectors in $\real^D$. This replaces the integrals in divergences and expectations with sums. Moreover, the latent prior $f_{\bb{Z}}$ is taken to be uniform over $\mathcal{Z}$ while the variational posterior $q$ for $\bb{Z}$ is taken to be a point estimate $\bb{z}_q\in\mathcal{Z}$.
The VQ-VAE encoder is realised as a function $\bb{z}_{e}\left(\xseq;\,\bb{\varphi}\right)$ taking values on all of $\real^D$, which subsequently is vector quantised using the nearest codebook vector to obtain $\bb{z}_q$.
After adding squared-error regularisation terms to the ELBO to promote codebook vectors and encoded values being close together, the full VQ-VAE objective function for a single datapoint becomes%
\footnote{This formula corrects a sign inconsistency present in Eq.\ (3) of \cite{van2017neural}.}
\begin{multline}
\cL_{\mathrm{VQ}}\given{\bb{\theta},\,\bb{\varphi},\,\mathcal{Z}}{\xseq}
= \ln f_{\Xseq\mid{}\bb{Z}}\givenpar{\xseq}{\bb{z}_q\left(\xseq\right)}{\bb{\theta}} \\
- \left\Vert \mathrm{sg}\left(\bb{z}_{e}\left(\xseq\right)\right)-\bb{z}_q\right\Vert_{2}^{2}
- \beta\left\Vert \bb{z}_{e}\left(\xseq\right)-\mathrm{sg}\left(\bb{z}_q\right)\right\Vert_{2}^{2}
\label{eq:vqvae}
\text{.}
\end{multline}
Here $\mathrm{sg}(\cdot)$ is the \emph{stop-gradient operator} implemented in many deep learning frameworks, which essentially means that the argument is to be treated as a constant during differentiation.
(For simplicity, we ignore the conditioning on $\lseq$ in our treatment of VQ-VAEs.)
The straight-through estimator described in \cite{bengio2013estimating} is used to backpropagate the gradient through the (non-differentiable) quantisation that turns $\bb{z}_e\left(\xseq\right)$ into $\bb{z}_q\left(\xseq\right)$ in the likelihood term. Since this estimator ignores the effect of the VQ codebook, the gradient used to update $\mathcal{Z}$ only depends on the second term in the objective function in Eq.\ \eqref{eq:vqvae} \cite{van2017neural}.

As originally introduced in \cite{van2017neural}, the regularisation terms in Eq.\ \eqref{eq:vqvae} (e.g., the ``commitment loss'') are motivated on geometric, not probabilistic grounds. Together with the quantisation and the stop-gradient operators, this makes it difficult to assign a probabilistic interpretation to the VQ-VAE objective function.
However, we will now show that it is possible to interpret the objective function as an actual ELBO maximisation.

\textbf{Proposition 1:} For $\beta=1$, optimising the VQ-VAE objective in Eq.\ \eqref{eq:vqvae} is equivalent to optimising the combined objective
\begin{multline}
\cL_{\mathrm{VQ1}}\given{\bb{\theta},\,\bb{\varphi},\,\mathcal{Z}}{\xseq} \\
= \ln f_{\Xseq\mid{}\bb{Z}}\givenpar{\xseq}{\bb{z}_q\left(\xseq\right)}{\bb{\theta}}
- \left\Vert \bb{z}_{e}\left(\xseq;\,\bb{\varphi}\right)-\bb{z}_q\right\Vert_{2}^{2}
\label{eq:vqvaecombo}
\text{,}
\end{multline}
which lacks the stop-gradient operators.


This proposition is easily verified by computing and comparing the partial derivatives of $\cL_{\mathrm{VQ}}$ and $\cL_{\mathrm{VQ1}}$ with respect to $\bb{\theta}$, $\bb{\varphi}$, and $\mathcal{Z}$. In practice, the results of learning are said \cite{van2017neural} not to depend substantially on the numerical value of the hyperparameter $\beta$. Our analysis will henceforth assume $\beta=1$, although $\beta=0.25$ is used for the experiments, following \cite{van2017neural}.

Next we will show how Eq.\ \eqref{eq:vqvaecombo} can be derived in a principled manner from a probabilistic model that includes a statistical model of the effect of quantisation in the latent space. We are not aware of any prior publications that derive VQ-VAEs from probabilistic principles alone.

To begin with, we model the distribution of encoder outputs in the latent space through a Gaussian mixture model (GMM). More concretely, we separate encoding and quantisation through a two-part latent variable $\bb{Z}=(\bb{Z}_e,\,\bb{Z}_q)$, where $\bb{z}_e\in\real^D$ represents the encoder output and $\bb{z}_q\in\mathcal{Z}\subset\real^D$ is the quantised version thereof.
Assume that $\Xseq$ is conditionally independent of $\bb{Z}_e$ given the codebook vector $\bb{Z}_q$. (This is the reverse of more conventional uses of mixture models in VAEs \cite{nalisnick2016approximate,tomczak2017vae}, where the observation $\Xseq$ is instead assumed to be conditionally independent of the mixture component identity $\bb{Z}_q$ given the mixture model sample $\bb{Z}_e$.) The joint model then factorises as
\begin{multline}
f_{\Xseq,\,\bb{Z}_e,\,\bb{Z}_q}\left(\xseq,\,\bb{z}_e,\,\bb{z}_q;\,\bb{\theta}\right) \\
= f_{\Xseq\mid{}\bb{Z}_q}\givenpar{\xseq}{\bb{z}_q}{\bb{\theta}}
f_{\bb{Z}_e\mid{}\bb{Z}_q}\given{\bb{z}_e}{\bb{z}_q}
f_{\bb{Z}_q}\left(\bb{z}_q\right)
\label{eq:vqvaefactor}
\text{.}
\end{multline}
We further assume that the latent prior $f_{\bb{Z}_q}$ over codebook vectors is uniform and that $f_{\bb{Z}_e\mid{}\bb{Z}_q}$ is an isotropic Gaussian centred on $\bb{Z}_q$ with fixed covariance matrix $\sigma^2\bb{I}$. $\bb{Z}_e$ here provides an explicit representation of the noise introduced by the vector quantiser. Analogous to a regular VAE, the remaining parameters $\bb{\varphi}$ and (here) $\mathcal{Z}$ define the variational posterior $q_{\bb{Z}}$. In particular, we choose a posterior of the form
\begin{align}
& \hphantom{=} q_{\bb{Z}_e,\,\bb{Z}_q\mid{}\Xseq} \givenpar{\bb{z}_e,\,\bb{z}_q}{\xseq}{\bb{\varphi},\,\bb{e}} \nonumber \\
& = q_{\bb{Z}_e\mid{}\bb{Z}_q,\,\Xseq}\givenpar{\bb{z}_e}{\bb{z}_q,\,\xseq}{\bb{\varphi}}
q_{\bb{Z}_q\mid{}\Xseq}\givenpar{\bb{z}_q}{\xseq}{\bb{e}} \\
& = f\left(\bb{z}_e - \bb{z}\left(\xseq;\,\bb{\varphi}\right)\right)
I\left(\bb{z}_q = \bb{e}\right)
\label{eq:vqvaeq}
\text{,}
\end{align}
Here, $\bb{e}\in\mathcal{Z}$ (to enforce quantisation), $I(\cdot)$ is the indicator distribution (which equals one if the argument is true and zero otherwise), while $f(\cdot)$ is any fixed, unimodal distribution centred on the origin. To reduce confusion with the latent outcome $\bb{z}_e$, we have abbreviated the encoder output $\bb{z}_{e}\left(\xseq;\,\bb{\varphi}\right)$ as $\bb{z}\left(\xseq;\,\bb{\varphi}\right)$.
When $f(\cdot)$ shrinks to a point mass, meaning that we ignore the uncertainty in the latent posterior, we call this model a \emph{GMM-quantised VAE}, or GMMQ-VAE.

\textbf{Proposition 2:} Under the assumptions made in \cite{van2017neural}, ELBO maximisation over the extended parameter set $\bb{\psi}=\left\{\bb{\theta},\,\bb{\varphi},\,\mathcal{Z},\,\bb{e}\in\mathcal{Z}\right\}$ for the GMMQ-VAE has the same form as parameter estimation with the VQ-VAE objective in Eq.\ \eqref{eq:vqvaecombo}.

\textbf{Proof sketch:} From Eq.\ \eqref{eq:elbo}, the GMMQ-VAE ELBO is
\begin{multline}
\elbo_{\mathrm{GMMQ}}\given{\bb{\psi}}{\xseq}
= -h\left(q_{\bb{Z}}\right) \\
+ \sum_{\bb{z}_q} \int q_{\bb{Z}}\left(\bb{z};\,\bb{\varphi},\,\bb{e}\right)
\ln f_{\Xseq,\,\bb{Z}}\left(\xseq,\,\bb{z};\,\bb{\theta}\right) \dif{}\bb{z}_e
\text{,}
\end{multline}
where $h(\cdot)$ denotes the differential entropy. Since the entropy of $q_{\bb{Z}}$ is independent of $\bb{\psi}$ it has no effect on ELBO maximisation and can be ignored. If we then let $f(\cdot)$ approach a Dirac delta function $\delta(\cdot)$ -- thus ignoring any uncertainty in the variational posterior by shrinking it to a point mass -- the sum and integral both reduce to simple evaluation, and we obtain
\begin{align}
& \estpsi
= \argmax_{\bb{\psi}} \lim_{f\to\delta} \elbo_{\mathrm{GMMQ}}\given{\bb{\psi}}{\xseq} \\
& = \argmax_{\bb{\psi}} \ln f_{\Xseq,\,\bb{Z}_e,\,\bb{Z}_q}\left(\xseq,\,\bb{z}\left(\xseq;\,\bb{\varphi}\right),\,\bb{e};\,\bb{\theta}\right) \\
& = \argmax_{\bb{\psi}} \left(
\ln f_{\Xseq\mid{}\bb{Z}_q}\givenpar{\xseq}{\bb{e}}{\bb{\theta}}
+ \ln f_{\bb{Z}_e\mid{}\bb{Z}_q}\given{\bb{z}\left(\xseq;\,\bb{\varphi}\right)}{\bb{e}} \right)
\text{,}
\end{align}
using Eq.\ \eqref{eq:vqvaefactor} with $f_{\bb{Z}_q}$ uniform. For the optimisation over $\bb{e}\in\mathcal{Z}$ in $\bb{\psi}$, $f_{\bb{Z}_e\mid{}\bb{Z}_q}$ is unimodal isotropic, and thus maximised by the $\bb{e}$ closest to $\bb{z}\left(\xseq;\,\bb{\varphi}\right)$. Also, for good autoencoders (i.e., near the global optimum of $\bb{\psi}\setminus{}\bb{e}$) we expect $f_{\Xseq\mid{}\bb{Z}_q}\givenpar{\xseq}{\bb{e}}{\bb{\theta}}$ to be greatest for the $\bb{e}\in\mathcal{Z}$ closest to $\bb{z}\left(\xseq;\,\bb{\varphi}\right)$. This is essentially a less restrictive version of the VQ-VAE assumption $f_{\Xseq\mid{}\bb{Z}_q}\givenpar{\xseq}{\bb{z}}{\bb{\theta}}\approx{}0$ whenever $\bb{z}\neq{}\bb{z}_q$ \cite{van2017neural}.
The optimisation over $\bb{e}$ can then solved explicitly, with the optimum being
\begin{align}
\bb{e}^{\star}
& = \bb{z}_q\left(\xseq;\,\bb{\varphi},\,\mathcal{Z}\right) \\
& = \argmin_{\bb{e}\in\mathcal{Z}}
\left\Vert \bb{z}\left(\xseq;\,\bb{\varphi}\right)-\bb{e}\right\Vert_{2}^{2}
\text{,}
\end{align}
the codebook vector closest to the encoder output $\bb{z}\left(\xseq;\,\bb{\varphi}\right)$, as expected for a vector quantiser. Since $f_{\bb{Z}_e\mid{}\bb{Z}_q}$ is Gaussian with covariance matrix $\sigma^2\bb{I}$, its log-probability reduces to the squared distance between the quantised and unquantised encoder output, plus a constant. We then arrive at
\begin{multline}
\left\{\esttheta,\,\estphi,\,\estZ\right\}
= \argmax_{\bb{\theta},\,\bb{\varphi},\,\mathcal{Z}} \bigg(
\ln f_{\Xseq\mid{}\bb{Z}_q}\givenpar{\xseq}{\bb{z}_q\left(\xseq;\,\bb{\varphi},\,\mathcal{Z}\right)}{\bb{\theta}} \\
- \frac{1}{2\sigma^2}\left\Vert \bb{z}\left(\xseq;\,\bb{\varphi}\right)-\bb{z}_q\left(\xseq;\,\bb{\varphi},\,\mathcal{Z}\right)\right\Vert_{2}^{2} \bigg)
\text{.}
\label{eq_probvqvae}
\end{multline}
This expression is of the same form as Eq.\ \eqref{eq:vqvaecombo}, as desired.
The variance $\sigma^2$ of the isotropic Gaussian acts as a weight between the two terms in the objective function, very similar to the hyperparameter $\beta$ in regular VQ-VAEs.


Proposition 2 shows that the entire VQ-VAE objective function for $\beta=1$ can be assigned a probabilistic interpretation as a regular VAE with a Gaussian mixture distribution in the latent space, specifically a GMMQ-VAE. The key twist is that $\Xseq$ depends on the discrete GMM component $\bb{z}_q$ instead of the continuous-valued, GMM-distributed encoder output $\bb{z}_e$ like in \cite{nalisnick2016approximate,tomczak2017vae}. This introduces quantisation into the encoder, distinguishing VQ-VAEs from the alternative of a simple, unquantised VAE with a GMM prior on $\bb{Z}$. We see that different weights on the squared-error term (which is closely related to changing $\beta$ in Eq.\ \eqref{eq:vqvae}) correspond to different assumptions about the magnitude of the quantisation error.

Our derivation of Proposition 2 suggests a number of natural generalisations of GMMQ/VQ-VAEs, for example by adjusting and potentially learning any combination of the component prior probabilities $f_{\bb{Z}_q}$ and the component covariance matrices $\bb{\Sigma}_q$. These extensions are however beyond the scope of the current article, and will not be explored further here. Since the GMMQ-VAEs and VQ-VAEs are so closely related, we will henceforth concentrate on VQ-VAEs for simplicity.


\subsection{A Variational Interpretation of Heuristic Control Learning}
\label{ssec:heuristics}
In this section, we show how discriminant condition codes \cite{abdel2013fast,xue2014fast,luong2017adapting,arik2017deep,taigman2018voiceloop} and sentence-level control vectors \cite{watts2015sentence}, which we collectively will refer to as the \emph{heuristic approaches} or \emph{poor man's latent variables}, can be connected to variational inference, autoencoders, and VQ-VAEs.
We begin by noting that the heuristic approaches are merely different names for the same model-fitting framework, where the likelihood maximisation in Eq.\ \eqref{eq:loglike} is replaced by a joint log-probability optimisation over both model parameters $\bb{\theta}$ and the per-sequence latent variables $\left\{\bb{z}^{(n)}\right\}$. The resulting estimation problem over the entire training data $\data$ can be written
\begin{multline}
\left\{\esttheta_{\mathrm{DCC}}(\data),\,\estz_{\mathrm{DCC}}^{(n)}(\data)\right\}
\\
= \argmax_{\left\{\bb{\theta},\bb{z}^{(n)}\right\}} 
\sum_{n=1}^{N} \ln f_{\Xseq\mid{}\bb{Z},\,\Lseq}\givenpar{\xseq^{(n)}}{\bb{z}^{(n)},\,\lseq^{(n)}}{\bb{\theta}}
\label{eq:heuristic}
\text{.}
\end{multline}

\textbf{Proposition 3:} The heuristic methods based on joint optimisation of latent inputs and model parameters equivalently be formulated encoder-decoder models, where the encoder for any $\bb{\theta}$ can be written  
\begin{multline}
\estz_{\mathrm{DCC}}^{(n)}(\data,\,\bb{\theta})
= \argmax_{\bb{z}} 
\ln f_{\Xseq\mid{}\bb{Z},\,\Lseq}\givenpar{\xseq^{(n)}}{\bb{z},\,\lseq^{(n)}}{\bb{\theta}}
\label{eq:encoder}
\text{.}
\end{multline}

\textbf{Proof sketch:} Consider 
\begin{align}
& \hphantom{=} \esttheta_{\mathrm{DCC}}(\data) \nonumber \\
& = \argmax_{\bb{\theta}}
\max_{\left\{\bb{z}^{(n)}\right\}}
\sum_{n=1}^{N} \ln f_{\Xseq\mid{}\bb{Z},\,\Lseq}\givenpar{\xseq^{(n)}}{\bb{z}^{(n)},\,\lseq^{(n)}}{\bb{\theta}} \\
& = \argmax_{\bb{\theta}}
\sum_{n=1}^{N} \max_{\bb{z}^{(n)}}
\ln f_{\Xseq\mid{}\bb{Z},\,\Lseq}\givenpar{\xseq^{(n)}}{\bb{z}^{(n)},\,\lseq^{(n)}}{\bb{\theta}} \\
& = \argmax_{\bb{\theta}}
\sum_{n=1}^{N}
\ln f_{\Xseq\mid{}\bb{Z},\,\Lseq}\givenpar{\xseq^{(n)}}{\estz_{\mathrm{DCC}}^{(n)}(\data,\,\bb{\theta}),\,\lseq^{(n)}}{\bb{\theta}}
\end{align}
where the last line follows from the observation that
\begin{align}
\max_{\bb{z}} g(\bb{x},\,\bb{z})
& = g(\bb{x},\,\argmax_{\bb{z}} g(\bb{x},\,\bb{z}))
\text{,}
\end{align}
for any function $g(\cdot,\,\cdot)$.

From Proposition 3 we observe that the common heuristics for learning controllable speech synthesis from unannotated data can be seen as encoder-decoder models, where the encoder uses the same network as the decoder. This observation motivates our interest in comparing these heuristics to other encoder-decoder approaches. (The situation is however different from traditional autoencoders with \emph{tied} weights, where the weight matrices in the decoder are transposes of those in the decoder.) Unlike VAEs, where encoding is performed via forward propagation through a second network, encoding here involves solving an optimisation problem through backpropagation. This is likely to be slow, but may give better performance (especially on test data) since each encoded variable solves an independent posterior-probability optimisation problem; there's no amortisation gap, unlike for VAEs \cite{shu2018amortized}. In both VAEs and in the heuristic framework the encoder requires $\xseq$ as well as $\lseq$ as input, and thus cannot easily be applied in situations where natural speech acoustics are unavailable.

Different from the style-token encoder in \cite{skerry2018towards,wang2018style} and the speaker encoder in \cite{jia2018transfer}, the encoder here has access to the text-derived features of the spoken utterance. This is likely to promote encoder output that is more complementary to the text (reduced redundancy), but may or may not be more transferable between different text prompts. Interestingly, while recent Tacotron and VoiceLoop publications \cite{skerry2018towards,wang2018style,nachmani2018fitting} have added explicit and distinct encoding networks similar to \mbox{(VQ-)VAEs}, previous work \cite{wang2017uncovering,taigman2018voiceloop} by these groups used backpropagation through the decoder as an implicit encoder, in the same way as the heuristic methods considered here.

\textbf{Proposition 4:} Increasing the heuristic objective function in Eq.\ \eqref{eq:heuristic} increases the evidence lower bound in Eq.\ \eqref{eq:elbo}. The encoder output can be seen as an approximate maximum a-posteriori estimate of the latent variable $\bb{Z}$ given $\Xseq=\xseq$ and $\Lseq=\lseq$.

\textbf{Proof sketch:} Note that the ELBO in Eq.\ \eqref{eq:elbo} can be written
\begin{multline}
\elbo\given{\bb{\theta},\,\bb{\varphi}}{\xseq,\,\lseq}
= \int q\left(\bb{z};\,\bb{\varphi}\right) \ln \frac{f_{\Xseq,\,\bb{Z}\mid{}\Lseq}\givenpar{\xseq,\,\bb{z}}{\lseq}{\bb{\theta}}}{q\left(\bb{z};\,\bb{\varphi}\right)} \dif{}\bb{z} \\
= \int q\left(\bb{z};\,\bb{\varphi}\right) \ln f_{\Xseq,\,\bb{Z}\mid{}\Lseq}\givenpar{\xseq,\,\bb{z}}{\lseq}{\bb{\theta}} \dif{}\bb{z}
- h(q)
\text{,}
\end{multline}
where $h(q)$ is the differential entropy of $q\left(\bb{z};\,\bb{\varphi}\right)$.
Consider choosing the $q$-distribution from a family which is parameterised by location $\bb{\mu}$ only, meaning that $\bb{\varphi}=\bb{\mu}$ and
\begin{align}
q\left(\bb{z};\,\bb{\mu}\right)
& \to q\left(\bb{z}-\bb{\mu}\right)
\text{.}
\end{align}
This makes $h(q\left(\bb{z};\,\bb{\mu}\right))$ independent of $\bb{\mu}$, and we get
\begin{align}
& \hphantom{=} \estmu \left(\xseq,\,\lseq,\, \bb{\theta}\right) \notag \\
& = \argmax_{\bb{\mu}}
\elbo\given{\bb{\theta},\,\bb{\mu}}{\xseq,\,\lseq} \\
& = \argmax_{\bb{\mu}}
\int q\left(\bb{z};\,\bb{\mu}\right) \ln f_{\Xseq,\,\bb{Z}\mid{}\Lseq}\givenpar{\xseq,\,\bb{z}}{\lseq}{\bb{\theta}} \dif{}\bb{z}
\text{.}
\end{align}
If the shape of the $q$-distribution(s) is made increasingly narrow (by making the variance tend to zero) so that it approaches a Dirac delta function $\delta(\cdot)$ we obtain
\begin{align}
& \hphantom{=} \lim_{q\to\delta} \estmu \left(\xseq,\,\lseq,\, \bb{\theta}\right) \notag \\
& = \argmax_{\bb{\mu}}
\int \delta\left(\bb{z}-\bb{\mu}\right) \ln f_{\Xseq,\,\bb{Z}\mid{}\Lseq}\givenpar{\xseq,\,\bb{z}}{\lseq}{\bb{\theta}} \dif{}\bb{z} \\
& = \argmax_{\bb{\mu}}
\ln f_{\Xseq,\,\bb{Z}\mid{}\Lseq}\givenpar{\xseq,\,\bb{\mu}}{\lseq}{\bb{\theta}} \\
& = \argmax_{\bb{\mu}}
\ln \left(
f_{\Xseq\mid{}\bb{Z},\,\Lseq}\givenpar{\xseq}{\bb{\mu},\,\lseq}{\bb{\theta}}
\cdot f_{\bb{Z}\mid{}\Lseq}\givenpar{\bb{\mu}}{\lseq}{\bb{\theta}} \right) \\
& = \argmax_{\bb{\mu}}
\ln f_{\Xseq\mid{}\bb{Z},\,\Lseq}\givenpar{\xseq}{\bb{\mu},\,\lseq}{\bb{\theta}}
\text{,}
\end{align}
where the last line assumes that $f_{\bb{Z}\mid{}\Lseq}$ is constant. By applying these approximations to each training datapoint independently one obtains Eq.\ \eqref{eq:heuristic}.

In summary, we have shown that the heuristic objective in Eq.\ \eqref{eq:heuristic} can be derived from variational principles assuming:
\begin{enumerate}
    \item That the prior distribution $f_{\bb{Z}\mid{}\Lseq}$ is flat (constant) across the range of $\bb{z}$- and $\lseq$-values considered.
    \item We use a Dirac delta function (a spike) to represent all latent posterior distributions.
    \label{it:map}
\end{enumerate}
Both assumptions are directly analogous to assumptions made in the probabilistic derivation of VQ-VAEs in Proposition 2: VQ-VAEs use a uniform prior over codebook vectors and do not represent any uncertainty in the (encoded) latents. This is another motivation for us to compare the heuristic approach to the largely similar functionality offered by VQ-VAEs.
The second assumption explains the nickname ``poor man's latent variables'', since we see that the heuristic objective does not afford any representation of uncertainty in the latent space.

If the listed assumptions are violated, the variational approximation need not produce a maximum of the true likelihood, though the agreement between the two methods is likely to be greater the more accurate the two assumptions are.
Unlike the EM-based derivation in \cite{henter2017principles}, the derivation presented here establishes that any simultaneous modification of that increases Eq.\ \eqref{eq:heuristic} also increases the likelihood lower bound; it is not necessary to perform interleaved optimisation as in the EM-algorithm \cite{dempster1977maximum}.

While $\elbo$ diverges to minus infinity as $q\to\delta$, and thus does not provide a reasonable numeric lower bound on the likelihood, it is still true that relative differences is $\elbo$-are meaningful and can be mapped to similar changes in the lower bound (consider subtracting one ELBO from another). A similar observation applies to the numerical value of the VQ-VAE objective derived in Proposition 2.

The domain of the optimisation over $\bb{z}^{(n)}$ in Eq.\ \eqref{eq:heuristic} can also be given a statistical interpretation. Define a binary prior $f_{\bb{Z}\mid{}\Lseq}$, which is constant and nonzero on feasible $\bb{z}$-values, but equals zero (so that $\ln f_{\bb{Z}\mid{}\Lseq}=-\infty$) outside the domain of optimisation. Unconstrained ELBO maximisation with this prior will then only find possible optimal parameters in the feasible set defined by the constraints. Constrained optimisation in the latent space is thus interpretable as normal variational parameter estimation under a particular prior on $\bb{Z}$.

To summarise, the key similarities between VQ-VAEs and the heuristic approach are:
\begin{itemize}
\item Both VQ-VAEs and the heuristic approach can be viewed as autoencoders.
\item Both methods are closely related to variational approaches with a flat prior over the permissible $\bb{z}$-values.
\item Neither approach represents uncertainty in the latent-variable inference (the encoder output value).
\end{itemize}
The main differences, meanwhile, are:
\begin{itemize}
\item The heuristic approach does not quantise latent vectors.
\item The heuristic approach uses a single network for both encoding and decoding, with an optimisation operation instead of forward propagation through a separate encoder. In other words, it does not amortise inference.
\end{itemize}

\subsection{Using Prior Information in Control Learning}
\label{ssec:prior}
It is worth noting that the variational interpretation of the heuristic method requires that a flat, noninformative prior is used. In Bayesian statistics, priors like $f_{\bb{Z}\mid{}\Lseq}$ can be adjusted by practitioners based on side information about what $\bb{z}$-value to expect for any given datapoint. With a fixed prior, this opportunity goes away.

There are, however, other methods for potentially biasing learning based on side information. In particular, since speech synthesisers are trained by local refinements of a previous parameter estimate and the parameter set includes explicit estimates of the latent encodings, the system can be initialised based on an informed guess about appropriate latent-variable values. We compare this strategy against random initalisation in the experiments in Sec.\ \ref{ssec:objective}. A finding that these two schemes do not differ in behaviour would indicate that learning is robust to initialisation. The opposite finding would suggest a more brittle learning process, but also one with room to straightforwardly inject prior information into the learning.

\section{Experiments}
\label{sec:experiments}
Following the theoretical developments in the previous section, we now investigate the practical performance of different methods for unsupervised learning of control in an example application to acoustic modelling of emotional speech, using a corpus described in Sec.\ \ref{ssec:data}. The systems and baselines considered are introduced in Sec.\ \ref{ssec:systems}, and their training presented in Sec.\ \ref{ssec:training}. The results of training and the associated learned latent representations are evaluated objectively in Sec.\ \ref{ssec:objective}. Sec.\ \ref{ssec:subjective} then details the subjective listening test performed, along with its analysis and resulting findings. Wherever possible, the experiments have been designed to be as similar as possible to the experiments with supervised speech-synthesis control in \cite{lorenzo2018investigating}, which used the same data.


\subsection{Data and Preprocessing}
\label{ssec:data}
For the experiments in this paper, we decided to use the large database of studio-recorded, high-quality acted emotional speech from \cite{lorenzo2018investigating}. (An earlier subset of this database was used for the research in \cite{henter2017principles}.)
The database contains recordings of isolated utterances in Japanese, read aloud by a female voice talent who is a native speaker of Japanese. Each prompt text was chosen to not harbour any inherent emotion, but was spoken in one or more of seven different emotional styles: emotionally-neutral speech as well as the three pairs happy vs.\ sad, calm vs.\ insecure, and excited vs.\ angry.
This means that the database contains speech variation of communicative importance that cannot be predicted from the text alone.
1200 utterances (133--158 min) were recorded for each emotion, for a total of 8400 utterances and nearly 17 hours of audio (beginning and ending silences included), all recorded at 48 kHz.
The talker was instructed to keep their expression of each emotion constant throughout the recordings.

Each audio recording in the data is annotated with the text prompt (in kanji and kana) as well as the prompted emotion. Lorenzo-Trueba et al.\ \cite{lorenzo2018investigating} considered a number of different methods for encoding this emotional information for speech synthesiser control, while also leveraging information on listener perception of the different emotions. They found the best-performing encoding of emotional categories to be based on listener responses to emotional speech (confusion-matrix columns) rather than one-hot categorical vectors. Re-labelling the data based on listener perception of individual utterances did not improve performance.
In contrast to this previous work, we will treat the emotional content as a latent source of variation, to be discovered and described by the different unsupervised methods we are investigating.

To simplify comparison, we used the same partitioning, preprocessing, and forced alignment of the database as Lorentzo-Trueba et al.\ \cite{lorenzo2018investigating}. In particular 10\% of the data were used for validation and 10\% for testing, with these held-out sets only incorporating sentences where annotators' perceived emotional categories agreed with the prompted emotion.
We also used the exact same linguistic and acoustic features as those extracted in \cite{lorenzo2018investigating}. In particular, Open JTalk \cite{oura2010japanese} was used to extract 389 linguistic features while WORLD \cite{morise2015cheaptrick,morise2016world} was used for acoustic analysis and signal synthesis. The analysis produced a total of 259 acoustic features at 5 ms intervals. The features comprised linearly interpolated log pitch estimated using SWIPE \cite{camacho2008sawtooth}, 60 mel-cepstrum features (MCEPs, with frequency warping 0.77 to approximate the Bark scale), and 25 band-aperiodicity coefficients (BAPs) based on critical bands. Each of these had static, delta, and delta-delta coefficients. These continuous-valued features were all normalised to zero mean and unit variance, and subsequently complemented with a binary voiced/unvoiced flag.

Linguistic and acoustic features were forced-aligned with five-state left-to-right no-skip HMMs trained with HTS \cite{zen2007hmm}, given access to the prompted emotion as an additional decision-tree feature. These HMMs were also used for duration prediction during synthesis, which was identical for all models; only different approaches to acoustic modelling (trained with or without emotional labels) were compared in the experiments.
At synthesis time, predicted static and dynamic features were reconciled through most likely parameter generation (MLPG) \cite{tokuda2000speech} and enhanced using the postfilter described in \cite{yoshimura2005incorporating} with coefficient 0.2.

\subsection{Systems}
\label{ssec:systems}
To investigate how supervised and unsupervised approaches for learning acoustic-model control behave on data with important non-textual variation (specifically emotion), we considered eight different sources of speech stimuli, or \emph{systems}, of three different kinds: stimuli based on natural speech (functioning as toplines), systems with only supervised learning (functioning as baselines for comparisons), and systems capable of learning output control from unannotated variation. In brief, the eight systems were defined as follows:
\begin{itemize}
\item \textbf{NAT:} Natural speech from the held-out test-set.
\item \textbf{VOC:} Natural speech from the held-out test-set, subjected to analysis synthesis as described in Sec.\ \ref{ssec:data}.
\item \textbf{SUP:} A supervised approach to controllable speech synthesis, trained and evaluated with labels derived from the ground-truth prompted emotion as input. Specifically, this system is equivalent to the best setup with emotional strength from \cite{lorenzo2018investigating}, since the approaches based on unannotated data presumably can learn to moderate emotional strength as well. The only difference from \cite{lorenzo2018investigating} is that the system was optimised using Adam \cite{kingma2015adam} rather than stochastic gradient descent.
\item \textbf{BOT:} A bottom-line system, same as SUP but with no control input, only linguistic features $\lseq$. This system cannot accommodate the differences between the different emotions in the database and provides a bottom line in terms of prediction performance.
\item \textbf{VQS:} A VQ-VAE with the same (`S') number of hidden nodes and layer order in the encoder as in the decoder.
\item \textbf{VQR:} A VQ-VAE with the same number of hidden nodes and but reverse (`R') layer order in the encoder compared to the decoder.
\item \textbf{HZI:} Poor man's latent variables with latent-space control vectors initialised with all zeros (`ZI').
\item \textbf{HSI:} Poor man's latent variables with supervised initialisation (`SI') of latent-space control vectors. 
This gives an idea of the impact of using prior information in initialisation, as discussed in Sec.\ \ref{ssec:prior}.
\end{itemize}

All synthesisers used the same duration model and duration predictions as the experiments in \cite{lorenzo2018investigating}; only the acoustic models differed. They also used exact same decoder structure, identical to the one used in \cite{wang2017autoregressive,henter2017principles,lorenzo2018investigating} (among others). Based on the proposal in \cite{fan2014tts}, it contains two 256-unit feedforward layers with logistic sigmoid nonlinearities, followed by two 128-unit BLSTM layers and a linear output layer. The neural networks were implemented in CURRENNT \cite{weninger2015introducing}.
\begin{figure}[!t]
\centerline{%
\subfloat[VQS (``same'')]{\includegraphics[width=0.8\columnwidth]{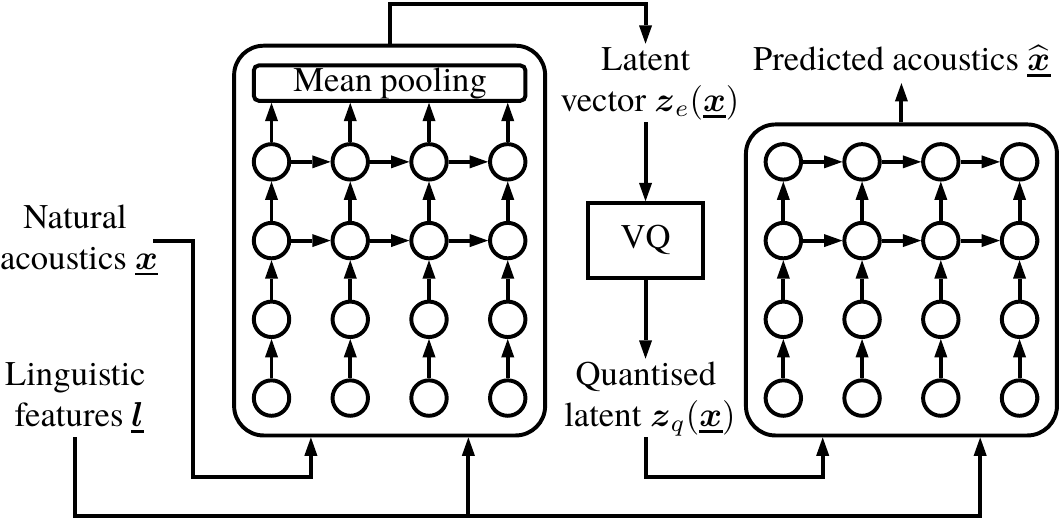}%
\label{sfig:vqvaesame}}}
\centerline{%
\subfloat[VQR (``reversed'')]{\includegraphics[width=0.8\columnwidth]{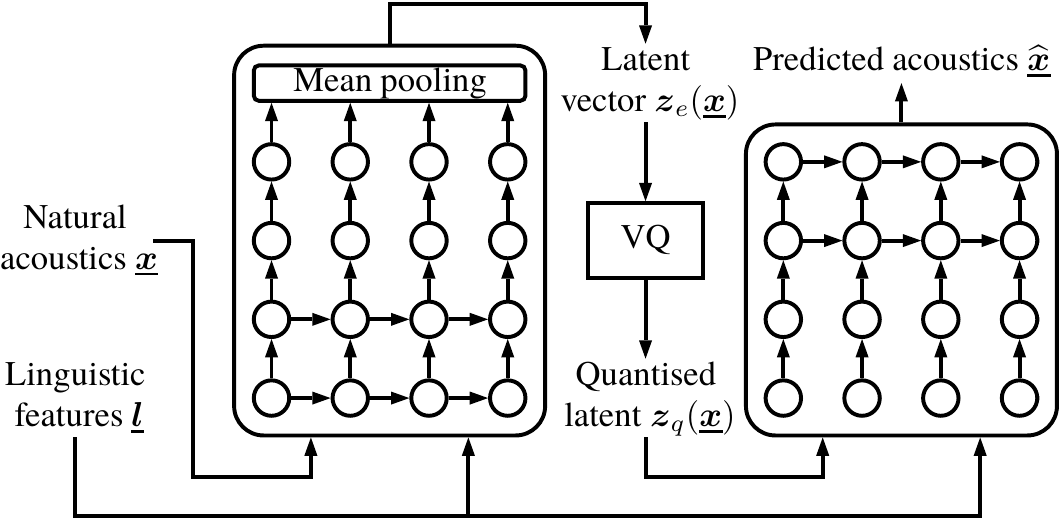}%
\label{sfig:vqvaereverse}}}
\caption{VQ-VAE schematics showing the two different encoder structures.}
\label{fig:vqvaestructure}
\end{figure}

Based on our observation in Prop.\ 3 in Sec.\ \ref{ssec:heuristics} -- that the heuristic methods can be interpreted as encoder-decoder models that use the same network for both encoding and decoding -- we made the VQ-VAE encoders in the experiments have the same internal structure (hidden layers and unit counts) as the decoder. There is, however, some ambiguity as for how to order the hidden layers in the encoder: the encoder is a function $\bb{z}_q\left(\xseq,\,\lseq\right)$ while the decoder is a function $\xseq\left(\bb{z}_q,\,\lseq\right)$. An argument based on $\bb{z}_q$ or $\xseq$ suggests that the order of the feedforward and recurrent layers be swapped in the encoder compared to the decoder, placing the recurrent layers closer to the input side of the encoder (as in system VQR), while a reference to $\lseq$ suggests that the layer order should not be altered between encoder and decoder (as in system VQS). The situation is illustrated in Fig.\ \ref{fig:vqvaestructure}. For completeness, both topologies were considered in the experiments. In either case, the final per-sentence encoding vector $\bb{z}_e$ was extracted from a mean-pooling layer across all timesteps, similar to how the backpropagated gradients for the latent control vectors sum across frames in the heuristic approach.

Prior to training, all networks were initialised with small random weights based on Glorot \& Bengio \cite{glorot2010understanding}. The autoencoder-based approaches in this study also require that the latent representations (the per-sentence control vectors or the codebook) be initialised as well. We set the control-vector dimensionality $D$ to 8 throughout the experiments, the same value as in \cite{lorenzo2018investigating} (based on 7 emotions plus a scalar emotional strength).
The latent control vector elements for HZI and HSI were then initialised deterministically (either all zeros, or with the same values as for as SUP, also on the validation and test sets).
For the VQ-VAEs the codebook size was set to 1344 and the codebook vectors were initialised with small random values as part of neural network initialisation.
The size of the codebook was chosen to be the same as the maximum number of distinct emotional-category encodings used by SUP on the training set \cite{lorenzo2018investigating}, computed as 192 35-utterance mini-batches with 7 emotions in each. It is good practice to use a larger VQ codebook than might be necessary, since some codebook vectors are likely to end up in regions that the encoder does not visit, yielding ``dead'' vectors that are neither trained or used; with too few vectors, the presence of local optima means that not all control modes or nuances may be learned. 
\begin{figure*}[!t]
\centerline{\subfloat[Training set]{\includegraphics[width=\columnwidth]{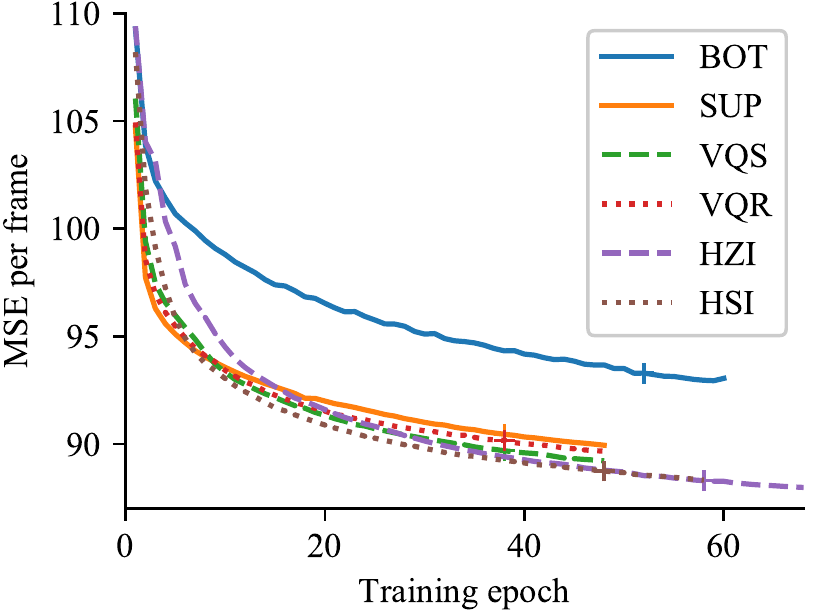}%
\label{sfig:traincurve}}%
\hfill%
\subfloat[Test set]{\includegraphics[width=\columnwidth]{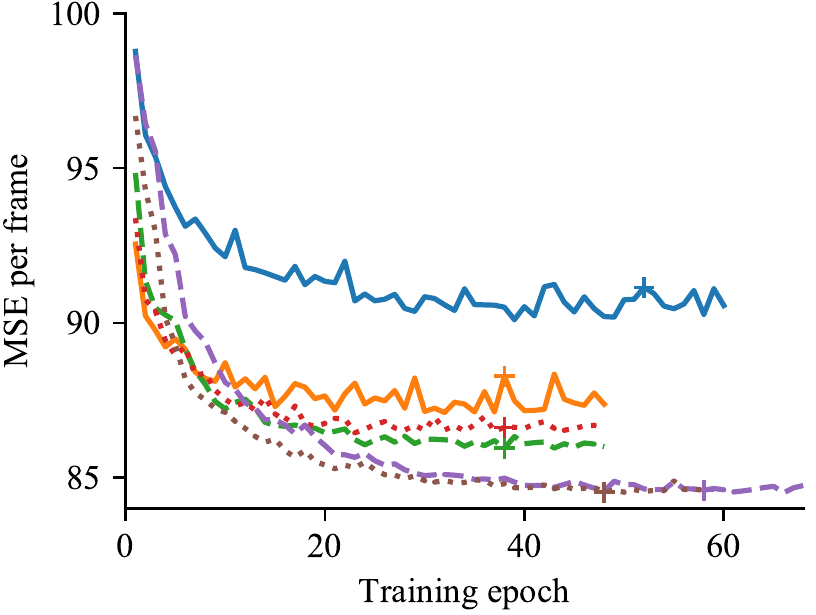}%
\label{sfig:testcurve}}}
\caption{Training curves for different systems. Note the different scales on the y-axes. Plus signs indicate the best epoch on the validation set.}
\label{fig:training}
\end{figure*}

In purely objective terms, we may expect the unsupervised approaches to achieve a better fit to the training data than the supervised method, since the former can tailor their output to each individual utterance in the corpus. The heuristic methods are furthermore likely to give better objective prediction accuracy than VQ-VAEs, due to the amortisation gap and the VQ-VAE restriction to a discrete set of latent-space values. Subjectively, however, SUP will be hard to beat, since it is trained using supervised knowledge to explicitly control the perceptually most relevant variation in the data.

\subsection{Training}
\label{ssec:training}
All mathematical approaches considered in this work are probabilistic methods that operate on the principle of likelihood maximisation. For this experiment, we assume that the conditional output distribution $\Xseq\left(\lseq,\,\bb{z}\right)$ (or $\Xseq\left(\lseq\right)$ for BOT) is an isotropic Gaussian with fixed variance. Log-likelihood maximisation is then mathematically equivalent to (mean) squared-error (MSE) minimisation. The MSE is a common loss function in synthesiser training, used for instance in Tacotron 1 and 2 \cite{wang2017tacotron,shen2018natural}. In our case each extracted acoustic feature is normalised to unit variance prior to neural network training (see \cite{lorenzo2018investigating}), so our setup altogether corresponds to an assumption that the speech-feature outputs are Gaussian, uncorrelated, and that each feature-vector element has a standard deviation proportional to the global standard deviation of that feature on the training set; the network outputs, in turn, can also be interpreted probabilistically as estimated conditional Gaussian means. It was seen in \cite{watts2016hmms} that the use of such a globally-constant covariance matrix did not significantly affect synthesis quality compared to the alternative of letting the variance depend on linguistic context.
\begin{table}[!t]
\caption{Objective results of system training.}
\label{tab:objective}
\centering
\begin{tabular}{|cccccc|}
\hline 
 & & & \multicolumn{3}{c|}{MSE per frame}\tabularnewline
System & \#NN weights & Best epoch & Train & Val. & Test\tabularnewline
\hline 
BOT & 1.58M & 52 & 93.3 & 105.1 & 91.1\tabularnewline
SUP & 1.58M & 38 & 90.5 & 101.3 & 88.3\tabularnewline
VQS & 3.24M & 38 & 89.7 & 100.2 & 86.0\tabularnewline
VQR & 3.18M & 38 & 90.2 & 100.7 & 86.6\tabularnewline
HZI & 1.58M & 58 & 88.3 & $\hphantom{\text{0}}$98.9 & 84.6\tabularnewline
HSI & 1.58M & 48 & 88.8 & $\hphantom{\text{0}}$98.9 & 84.5\tabularnewline
\hline 
\end{tabular}
\end{table}

Encoder and decoder parameters (including the VQ codebook) were trained to minimise per-frame MSE using Adam \cite{kingma2015adam} with default hyperparameter values. However, since each per-utterance control-vector input for the heuristic systems HZI and HSI only is updated once per epoch, these $\bb{z}$-vectors may not be a good fit for the per-parameter moment estimates that Adam maintains. The control vectors were therefore instead updated using stochastic gradient descent (SGD) with a fixed learning rate $2\cdot10^{-4}$, the same rate as used for the latent vectors in \cite{henter2017principles}.%
\footnote{Paper \cite{henter2017principles} contains a typo where $0.2\cdot10^{-3}$ is incorrectly listed as $2\cdot10^{-3}$.}
The HZI and HSI control-vector inputs for validation and test utterances were updated similarly using the corresponding synthesis network from each epoch, but without modifying the network weights on these utterances (cf.\ \cite{watts2015sentence}). In an encoder-decoder view, this maximisation performed by SGD on training, validation, and test data is an instantiation of the encoder in Eq.\ \eqref{eq:encoder}.

Training was run until the validation-set MSE failed to improve for ten consecutive epochs (or eight in the case of BOT), whereafter the network with the lowest validation-set error was returned. In the present experiment, this scheme required at most 68 epochs for termination.

\subsection{Objective Evaluation}
\label{ssec:objective}
\subsubsection{Evaluation of Training}
Fig.\ \ref{fig:training} presents learning curves from the synthetic systems in Sec.\ \ref{ssec:systems}, chronicling the evolution of per-frame mean-squared error on training and test-set data for each epoch of optimisation. The number of iterations until termination and final performance numbers on all three data partitions are listed in Table \ref{tab:objective}, along with the number of neural network weights used by CURRENT for each system.
\begin{figure*}[!t]
\centerline{\subfloat[SUP]{\includegraphics[width=\columnwidth]{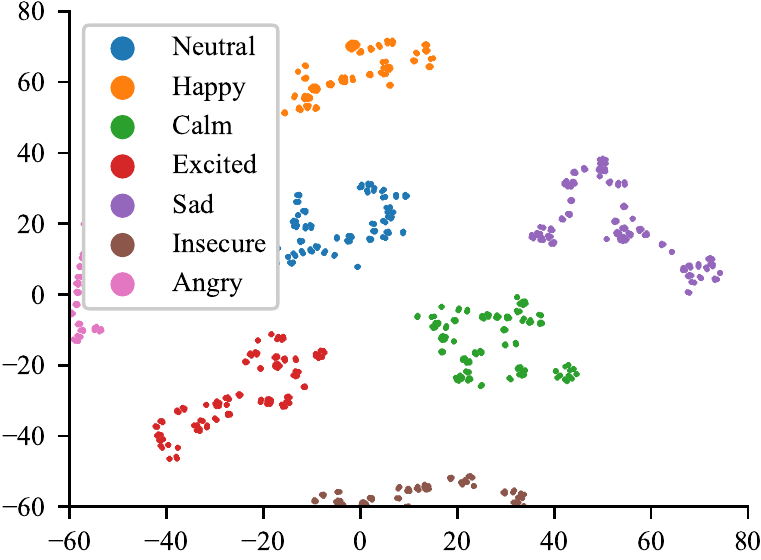}%
\label{sfig:supscatter}}%
\hfill%
\subfloat[HZI]{\includegraphics[width=\columnwidth]{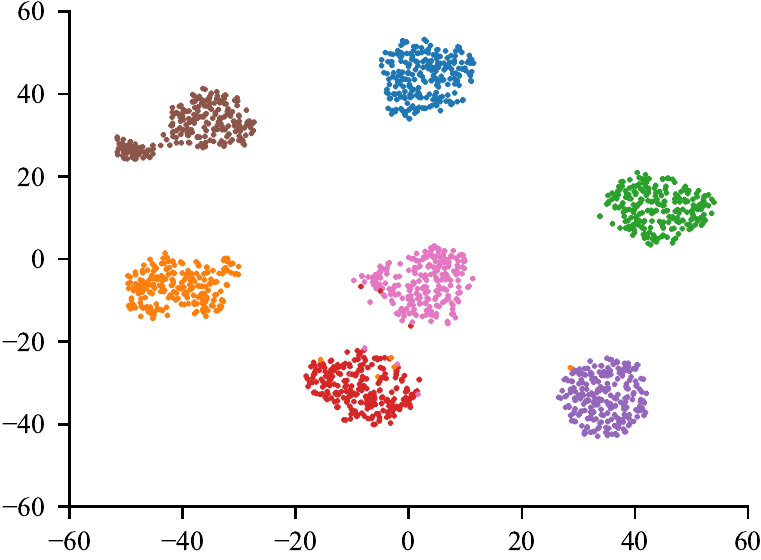}%
\label{sfig:hziscatter}}}
\caption{2D t-SNE embeddings of latent control vectors $\bb{z}$, coloured by the prompted emotion. Scale and rotation are arbitrary.}
\label{fig:scatter}
\end{figure*}

Looking at Table \ref{tab:objective}, a handful of general trends become evident. To begin with, validation set numbers are consistently inferior to both training and test set numbers; this appears to be a consequence of the data partitioning in \cite{lorenzo2018investigating}, and recurs in other systems trained on this data split.
The most notable difference between the methods is that all schemes with control achieved better MSE performance than the emotionally-unaware bottom line BOT by at least 3.0 on all data partitions. This is entirely expected, since only BOT is unable to adjust its output based on the emotional content of the speech.
The fact that methods with learned control inputs slightly outdo SUP is not surprising either, since they had access to the natural ground-truth acoustics for each test-set utterance as a decoder input. These numbers do not imply that the resulting systems achieve subjectively better quality or emotional control.

The heuristic systems required more epochs than most other systems to terminate training, but also achieved lower per-frame MSE than VQS and VQR by at least 1.4 on the test set. This difference is likely due to the amortisation gap \cite{cremer2018inference}, since the VQ-VAEs use learned inference while the heuristic systems use direct per-utterance optimisation.
The use of SGD rather than Adam for updating the latent-variable values of each utterance might explain the slower convergence rate and longer training seen in Fig.\ \ref{fig:training} for the heuristic systems.


As a side note, an earlier version of our VQ-VAE encoder extracted the final state on the LSTM (in each direction) and mapped these to the latent space through a linear output layer; such a design is perhaps more traditional in encoder-decoder models, and resembles the one used in \cite{wang2018style}. However, VQ-VAEs with this encoder design did not perform much differently from BOT. It seems that relevant information from mid-utterance acoustics did not propagate well to the end states, resulting in encoder output of little predictive value. Without emotional information (from label or acoustics), the resulting network is then essentially a version of BOT. Once the choice to extract the end state of the LSTM was replaced by a mean pooling operation, performance improved to the levels seen in Table \ref{tab:objective}.%
\footnote{As an alternative, the work in \cite{skerry2018towards} chose used the final state of a unidirectional RNN as the encoder output, but since their encoder contained several strided convolutions, the training sequences were effectively downsampled such that the RNN had to run over less than ten timesteps. Similar to our mean pooling, this allowed the encoder to better incorporate information from the entire utterance, but their setup is more likely to retain some order information of relevance to the intonation patterns they studied.}

\subsubsection{Evaluation of Learned Latent Vectors}
\label{sssec:latenteval}
While the low MSE achieved by the encoder-decoder models in Table\ \ref{tab:objective} are encouraging, it does not follow that the trained systems must have learned to represent and control emotion specifically. To investigate this, we performed objective analyses on the learned latent representations. For the heuristic systems, we used $t$-distributed stochastic neighbour embedding (t-SNE) \cite{maaten2008visualizing} to reduce dimensionality and visualise the latent-space vectors in two dimensions. The results for HZI can be seen in Fig.\ \ref{sfig:hziscatter}, and can be compared against a similar embedding of the SUP control vectors in Fig.\ \ref{sfig:supscatter}. It is clear that the different emotions are grouped into well-defined clusters with minimal overlap. The degree of separation can be quantified by looking at how frequently the nearest neighbour of an utterance vector in the latent space is from a different prompted emotion. Across the 1680 latent vectors in the test set, this happened 18 times for HZI and 7 times for HSI. If we measure how many times at least one of the five nearest neighbours is from a different emotion, the numbers rise to 41 for HZI and 21 for HSI. (For SUP, the corresponding number is 0.) All in all, this indicates that the heuristic approach has been highly successful at identifying the different base emotions in the database and then separating them in the latent space.

While exhibiting faster convergence, supervised initialisation (HSI) did not seem to confer any lasting benefit over the purely unsupervised approach HZI initialised with all zeros. This suggests that latent vectors learned through standard heuristics are robust against differences in initialisation.

For the systems based on VQ-VAE we performed a clustering analysis on the 1680 quantised latent vectors $\bb{z}_q$ from the test set. The results are provided in Table \ref{tab:vqclustering}. We see that most vectors in the codebooks were not used at all (at most 61 vectors out of 1344 were used), so a parsimonious discrete representation was learned despite starting from a very large codebook. Of the vectors that did see use on the test set, each emotion only used a subset of these (first group of numbers in the table). Standard measures of clustering quality like purity and normalised mutual information (NMI) \cite[Ch.\ 16]{manning2008introduction} indicate that the prompted emotions were very well separated by the VQ-VAE. Beyond the emotion, there is relatively little information in the encoded latent vectors, as shown by the low per-emotion entropies (second set of numbers in the table). This suggests that the talker's emotional expression might be quite consistent across the database, precisely as intended during recording, and does not leave much room for the encoded vectors $\bb{z}_q$ to pick up additional nuances in emotional expression. While VQR seems to yield smaller and more well-defined clusters than VQS, the differences are marginal and unlikely to have substantial impact on the synthesis.
\begin{table}[!t]
\caption{Analysis of quantised latent vectors in VQ-VAE systems.}
\label{tab:vqclustering}
\centering
{\setlength{\tabcolsep}{0.51em}%
\begin{tabular}{|c|c|c|ccc|}
\hline 
 & \multicolumn{1}{c|}{VQ indices used} & \multicolumn{1}{c|}{Emotion entropy} & Total & Purity & NMI\tabularnewline
System & min / mean / max & min / mean / max & indices & (frac) & (bits)\tabularnewline
\hline 
VQS & 2 / 11.7 / 33 & 0.19 / 2.03 / 3.98 & 61 & 0.96 & 0.17\tabularnewline
VQR & 1 / $\hphantom{\text{0}}$5.7 / 13 & 0$\hphantom{\text{.00}}$ / 1.24 / 2.71 & 29 & 0.98 & 0.10\tabularnewline
\hline 
\end{tabular}}
\end{table}
\begin{table}[!t]
\caption{Mean opinion scores for quality and emotional strength.}
\label{tab:mos}
\centering
\begin{tabular}{|c|cc|cc|}
\hline 
 & \multicolumn{2}{c|}{Quality} & \multicolumn{2}{c|}{Emotional strength}\tabularnewline
System & Per utt. & Per emo. & Per utt. & Per emo.\tabularnewline
\hline 
NAT & 4.01 & - & 3.38 & -\tabularnewline
VOC & 2.94 & - & 3.18 & -\tabularnewline
SUP & 3.41 & - & 2.94 & -\tabularnewline
\hline 
VQS & 3.42 & 3.51 & 2.92 & 2.99\tabularnewline
VQR & 3.41 & 3.50 & 2.89 & 2.97\tabularnewline
HZI & 3.43 & 3.53 & 2.89 & 2.99\tabularnewline
HSI & 3.44 & 3.54 & 2.86 & 2.98\tabularnewline
\hline 
\end{tabular}
\end{table}

In summary, we find that the unsupervised methods very successfully identified the emotional classes in held-out speech data on our task, despite not having access to explicit emotional annotation. This confirms that these methods are capable of identifying and representing salient, unannotated variation in the data, just like the unsupervised style tokens in \cite{wang2018style}.


\subsection{Subjective Evaluation}
\label{ssec:subjective}
Reduced objective error does not necessarily imply a perceptually better system. In fact, the true minimiser of the MSE objective we use is the conditional mean of $\Xseq$. This mean was estimated directly from repeated speech in \cite{henter2014measuring} and found to be perceptually inferior to random sampling in highly accurate models.
In order not to be led astray by the objective performance, we complemented our observations above with a crowdsourced subjective listening test similar to those in \cite{lorenzo2018investigating}.

\subsubsection{Listening Test Design}
For the listening test, the BOT system was excluded, as it is incapable of control. Each of the four unsupervised systems, however, was represented twice: 
once synthesising from control vectors derived from encoding the ground-truth held-out test sentences (the normal autoencoder approach), and once with the latent input to the encoder always set equal to the mean latent vector $\overline{\bb{z}}$ for the relevant emotion across the entire training set. While the former control scheme varies the control input $\bb{z}$ from utterance to utterance, the latter holds $\bb{z}$ constant for each emotion, wherefore we refer to these schemes as \emph{per-utterance} and \emph{per-emotion} control, respectively.

Our per-utterance control may in principle be able to reproduce nuances in the emotional expression of each test utterance, but requires access to the held-out test-set acoustics to do so.
Per-emotion control is derived from emotional labels on the training data (instead of using test-set acoustics), but any systematic variation in perceived emotional strength across utterances must then be attributed to the text input alone.
Together, the two control schemes can be used to assess the systems' abilities to replicate nuances in emotional expression on the test set.
Many other control schemes are also possible, but studying them is left as future work.

A system paired with a control scheme will be termed a \emph{condition}, of which we investigated a total of 11: NAT, VOC, SUP, and two each (for the two control schemes) for the unsupervised systems VQS, VQR, HZI, and HSI. Each of the 1680 utterances in the test set (240 per emotion) can then be realised in any condition, producing a \emph{stimulus} waveform.

Our subjective evaluation recruited native Japanese listeners through CrowdWorks\textsuperscript{LTD} to evaluate sets of 22 randomly-selected stimuli through a web-based interface. The sets were constrained such that all stimuli were unique and each condition appeared exactly twice in each set.
No listener was permitted to evaluate more than 10 sets.

Evaluators processed the stimuli in the set in sequence. For each stimulus, they were asked to supply three pieces of information: i) perceived speech quality (traditional MOS scale of integers ``1 -- bad'' through ``5 -- excellent''); ii) perceived emotional category (response options being the seven emotions in the database plus ``other''); and iii) perceived emotional strength (integer scale ``1 -- almost no emotion'' through ``5 -- very emotional'', or 6 for ``no emotion''). Evaluators could listen to each stimulus as many times as desired before responding. 
In total, 700 response triplets were gathered for each emotion, from a total of 50 different listeners.

\subsubsection{Evaluation of Synthesis Quality}
The first set of columns in Table \ref{tab:mos} shows the mean opinion scores (MOS) for speech quality for the different systems and control strategies investigated. To check if the differences were significant we applied two-sided Mann-Whitney U tests comparing all condition pairs, with Holm-Bonferroni correction \cite{holm1979simple} used to keep the familywise error rate below 5\%. These tests found NAT and VOC to be significantly different from all other systems, as well as from each other. No other differences in quality were found to be statistically significant. $t$-tests (also with Holm-Bonferroni correction) gave the same conclusions. We thus observe that SPSS, while not achieving the same performance as natural speech, can achieve good output quality both through supervised as well as unsupervised control in this application. The difference between the best and the worst (SUP) synthesiser MOS is a mere 0.13 points on the five-point MOS scale.
While there was evidence of a minor amortisation gap between VQ-VAEs and heuristic systems in terms of objective performance (i.e., MSE), this gap does not appear to have affected speech quality. Given that VQ-VAEs have advantages of being easier to train and allow straightforward latent-variable inference through amortisation, this makes them an appealing practical choice.

\subsubsection{Evaluation of Output Control}
Our primary interest in this work is not synthesis quality but controllability. We therefore assessed the synthesisers' ability to reproduce the emotions in the database by studying the emotional classifications assigned by the listeners in the listening test. These classifications can be summarised through a confusion matrix, tabulating the distribution of listener classifications conditioned on the different prompted emotions.
In the ideal case when all emotions are perceived as intended, this matrix should be the identity matrix. For completely natural speech there are nonetheless some confusions between emotions (as discussed in \cite{lorenzo2018investigating}), leading to some off-diagonal matrix structure.
\begin{table}[!t]
\caption{Frobenius distances between emotional confusion matrices. The best unsupervised performance in each column is bolded.}
\label{tab:confus}
\centering
\begin{tabular}{|c|ccc|ccc|}
\hline 
 & \multicolumn{3}{c|}{Per-utterance control} & \multicolumn{3}{c|}{Per-emotion control}\tabularnewline
System & vs.\ ID & vs.\ ref & vs.\ NAT & vs.\ ID & vs.\ ref & vs.\ NAT\tabularnewline
\hline 
NAT & 0.50 & 1.04 & 0.00 & - & - & -\tabularnewline
VOC & 0.68 & 1.26 & 0.37 & - & - & -\tabularnewline
SUP & 0.71 & 1.51 & 0.69 & - & - & -\tabularnewline
\hline 
VQS & 0.63 & 1.39 & \textbf{0.46} & \textbf{0.48} & \textbf{1.27} & \textbf{0.53}\tabularnewline
VQR & \textbf{0.58} & \textbf{1.35} & 0.51 & 0.65 & 1.44 & 0.70\tabularnewline
HZI & 0.60 & 1.39 & 0.53 & 0.59 & 1.37 & 0.55\tabularnewline
HSI & 0.64 & 1.42 & 0.52 & 0.62 & 1.42 & 0.63\tabularnewline
\hline 
\end{tabular}
\end{table}

Following the same methodology as in \cite[Sec.\ 8.1.1]{lorenzo2018investigating}, we computed emotional classification confusion matrices for each and every condition in the listening test (700 classifications per condition). These matrices were then compared against three different reference matrices: the ideal (identity matrix, `ID') as well as two confusion matrices from natural speech, namely the one tabulated in \cite[Table 5]{lorenzo2018investigating} (`ref') as well as the one computed from listener classifications of natural speech in the present listening test (`NAT'). Specifically, we computed the Frobenius norm of the difference between every confusion matrix and every reference matrix. Table \ref{tab:confus} presents the results of this comparison. A system that well separates and reproduces the different emotions should have low distance to the three references in the table.

While identifying statistically significant differences between confusion matrices is not a solved problem (see, e.g., \cite{leijon2016bayesian}), we note that (with one single exception) NAT is better than all other conditions in all metrics; this agrees with our expectation that the recorded natural speech should perform at least as well as SPSS control schemes learned from the same data. On the other end of the spectrum, SUP is found to have greater distance to the reference matrices than all other conditions (again with a single exception). All other conditions exhibit broadly comparable numbers for each reference. Taken together, these patterns suggest that unsupervised approaches are at least as good (or better) than supervised learning of control in the present application, but that there is little difference between VQ-VAEs and the heuristic methods (and between different control schemes) in how reliably they reproduce the base emotions in the corpus.

As the controllable speech synthesisers considered in this work are capable of control inputs that differentiate more than just the seven base emotions, there is the possibility that they may learn to control other aspects of speech variability such as emotional nuance (cf.\ \cite{henter2017principles}), assuming such variation is present in the training data. This might be reflected in the emotional strength ratings, whose means are tabulated in the last two columns of Table \ref{tab:mos}. (For this analysis, a response of ``no emotion'' was mapped to an emotional strength of zero.) Holm-Bonferroni corrected Mann-Whitney U tests between conditions (the same methodology used to analyse synthesis quality earlier) show that NAT and VOC perform similarly, and better than other conditions, which otherwise exhibit no significant differences. Thus the unsupervised approaches are again competitive with the supervised system.

No differences are evident between per-utterance and per-emotion control in this evaluation. This might not be too surprising, given the lack of diversity (only one or two bits of entropy) observed in Table \ref{tab:vqclustering} among control inputs in the same emotion class. Such a finding is consistent with expectations that the range of nuances within each emotion is quite limited in our speech corpus. It is possible that exaggerating the differences between utterance control inputs, as done in \cite{henter2017principles}, would give more noticeable differences in expression within each emotion class.


To summarise, we have found that the unsupervised approaches under consideration are comparable to the supervised system also in terms of perceived speech quality, emotion recognition, and perceived emotional strength. Moreover, the different unsupervised systems and control schemes appear essentially perceptually equivalent in our evaluation.

\section{Conclusion}
\label{sec:conclusion}
This paper has studied the theory and practice of unsupervised learning of output control in statistical text to speech. On the theory side, we have established novel connections between traditional unsupervised heuristics from speech-technology, like DCC and sentence-level control vectors, and variational latent-variable inference in autoencoder models.
We have likewise connected the heuristics to VQ-VAEs, which we have shown have a similar interpretation as variational inference neglecting uncertainty in a Gaussian mixture model.

In terms of empirical insights, we have compared supervised and unsupervised methods for learning controllable acoustic models on a large corpus of emotional speech. The objective and subjective results show that the unsupervised methods successfully learn and reproduce the emotional classes in the speech data and often outperform a competitive supervised baseline. This bodes well for unsupervised learning for enabling output control in speech synthesis at large. Methods incorporating amortised inference stand out as particularly appealing for future applications, since they achieve similar performance as the established heuristics but enable easier training and latent-variable inference.

\bibliographystyle{IEEEtran}
\bibliography{refs}


\end{document}